\documentclass[a4p,12pt,onecolumn,oneside,notitlepage,final]{article}
\usepackage{latexsym}
\usepackage{bm}
\usepackage{amsmath,amssymb}
\usepackage{mathrsfs}
\usepackage{tabularx}
\usepackage{float}
\usepackage{enumerate}
\usepackage[dvips]{graphicx}

\newcommand{\A}{\ddot{\rm{a}}}
\newcommand{\Pl}{\rm{Pl}}

\newcommand{\gaugino}{\lambda_a}
\begin{document}
\begin{flushright}
   TU-809 
\end{flushright}
\thispagestyle{empty}
\begin{center}
{\Large{\bf{Axionic Mirage Mediation}}} \\ 
\end{center}
\begin{center}
Shuntaro Nakamura, Ken-ichi Okumura and Masahiro Yamaguchi \\
\textit{Department of Physics, Tohoku University, Sendai, 980-8578, Japan}
\end{center}
\begin{center}
{\large{Abstract}}
\end{center} 
Although the mirage mediation is one of the most plausible mediation
mechanisms of supersymmetry breaking, it suffers from two crucial
problems. One is the $\mu$-/$B \mu$-problem and the second is the
cosmological one.
The former stems from the fact that the $B$ parameter tends to be
comparable with the gravitino mass, which is two order of magnitude
larger than the other soft masses. The latter problem is caused by the
decay of the modulus whose branching ratio into the gravitino pair is
sizable.
In this paper, we propose a model of mirage mediation,  in which
Peccei-Quinn symmetry is incorporated. In this \textit{axionic mirage
mediation}, it is shown that the Peccei-Quinn symmetry breaking scale is dynamically determined 
around $10^{10}$ GeV to $10^{12}$ GeV due to the supersymmetry breaking effects, and 
the $\mu$-problem can be solved
naturally. Furthermore, in our model, the lightest supersymmetric
particle (LSP) is the axino, that is the superpartner of the axion.
The overabundance of the LSPs due to decays of modulus/gravitino,
which is the most serious cosmological difficulty in the mirage
mediation, can be avoided if the axino is sufficiently light.  The
next-LSPs (NLSPs) produced by the gravitino decay eventually decay into
the axino LSPs, yielding the dominant component of the axinos
remaining today.  It is shown that the axino with the mass of 
$\mathcal{O}(100)$ MeV is naturally realized, which can constitute the dark matter of the
Universe, with the free-streaming length of the order of 0.1 Mpc. The
saxion, the real scalar component of the axion supermultiplet, can also
be cosmologically harmless due to the dilution of the modulus decay.
The lifetime of NLSP is relatively long, but much shorter than 1 sec., when the big-bang
nucleosynthesis commences. The decay of NLSP would provide intriguing collider signatures.

\newpage
\section{Introduction}
Supersymmetry (SUSY) is one of the most promising candidates for new physics beyond the standard model (SM), which solves the naturalness problem associated with the hierarchy between the electroweak scale and the Planck scale.  
However, since no supersymmetric particles have been discovered yet, SUSY has to be broken. 
In order to prevent us from arising the new dangerous phenomenological effects, flavor-changing-neutral-current (FCNC) process and SUSY CP violation and so on, the SUSY-breaking mediation mechanism must be the one that does not give rise to these effects. 
In superstring theory, one of most plausible mediation mechanisms without these effects is the moduli mediation \cite{Kaplunovsky:1993rd}. The contribution of the modulus, $X$, to the soft masses is proportional to its auxiliary component, $F_X \simeq m^2_{3/2}/m_X$ (in the Planck unit $M_{\Pl} = 1$), where $m_{3/2}$ and $m_X$ are the gravitino mass and the modulus mass, respectively. 
On the other hand, in 4-dimensional supergravity (SUGRA), the mediation associated with the super-Weyl anomaly, the anomaly mediation \cite{Randall:1998uk}--\cite{Bagger:1999rd}, also does exist, whose contribution to the soft masses is of the order of $m_{3/2}/8 \pi^2$.   
Recently, KKLT \cite{Kachru:2003aw} have proposed the interesting set-up to stabilize the modulus with the relatively heavy mass, $m_X \simeq 4 \pi^2 m_{3/2}$, by considering some non-perturbative effects. Therefore, in this and similar set-ups, the anomaly mediation contribution is comparable with the modulus mediation one. 
This type of mediation mechanism is often called the mirage mediation  \cite{Choi:2004sx}--\cite{Nagai:2007ud}. 

The mirage mediation is a natural mediation mechanism and quite interesting because it can solve the tachyonic slepton problem in the pure anomaly mediation, as well as has characteristic mass spectrums \cite{Endo:2005uy} \cite{Choi:2005uz}.
However, it has two crucial problems. One of them is the $\mu$-/$B \mu$-problem. In the discussion of ref.\cite{Choi:2005uz}, although it is not difficult to obtain that $\mu$ is of the order of the soft masses, $B$ becomes of the order of $m_{3/2}$, in general, without fine-tuning of parameters. In the mirage mediation, since the gravitino mass should be $\mathcal{O}(10)$ TeV in order to obtain soft masses with the electroweak scale, this is problematic. 

The other problem results from cosmology. While in inflation the modulus field is shifted from the true minimum. When the Hubble parameter $H$ decreased to the order of the modulus mass, $m_X$, the modulus starts to oscillate around the true minimum with amplitude of order the Planck scale. 
The energy density of the coherent oscillation dominates that of the Universe before the modulus decay. 
If the modulus mass is around the electroweak scale, its decay produces immense amount of entropy after primordial nucleosynthesis and upsets the success of the big-bang nucleosynthesis (BBN). Such a cosmological catastrophe is known as the moduli problem \cite{de Carlos:1993jw}. 
One of the resolutions to the moduli problem is to invoke a relatively heavy modulus with mass of $10^5$ GeV or larger. 
The heavy modulus then decays before the nucleosynthesis commences, which will not spoil it. 
Thus, KKLT set-up seems to be favored cosmologically.   

However, recently, it was found that non-thermal production of the gravitino from the modulus decay is sizable, which aggravates the moduli problem \cite{Endo:2006zj} \cite{Nakamura:2006uc}. (See, for the case of the inflaton and the Polonyi field \cite{Asaka:2006bv}--\cite{Nakamura:2007wr}). 
Such aggravation comes not only from abundance of the gravitino but also from the overclosure of the neutralinos  if it is the lightest superparticle (LSP). 
The gravitinos with the electroweak scale mass decay into minimal supersymmetric standard model (MSSM) particles after or during the primordial nucleosynthesis, and hence the decay products may spoil the success of BBN. This is called the gravitino problem. It is known that when the gravitino mass is as large as $\mathcal{O} (10)$ TeV, there is no gravitino problem because of their sufficiently short lifetime. 
On the other hand, the neutralinos produced by the decay of the gravitinos are so abundant that the annihilation process is effective. However, its abundance after the annihilation exceeds the present dark matter (DM) abundance even if the neutralino LSP is the wino, whose annihilation cross section is most effective.  
Therefore, even if the modulus can decay before BBN, it causes the cosmological problem. 
This disaster may be remedied when  LSP is  lighter than the neutralino.

We will solve these two problems simultaneously by the axionic extension of the mirage mediation (\textit{axionic mirage mediation}). 
In SM, the instanton effect of the quantum chromodynamics (QCD) generate the following term in the Lagrangian 
\begin{eqnarray}
   \mathcal{L}_{\bar{\Theta}} = \frac{g_3^2}{64 \pi^2} 
                         \bar{\Theta} \, \epsilon^{\mu \nu \rho \sigma} G^a_{\mu \nu} G^a_{\rho \sigma}, 
\end{eqnarray}
where $g_3$ is the strong coupling constant, $G^a_{\mu \nu}$ the field strength of the gluon and $\bar{\Theta} = \Theta + {\rm arg}( {\rm det} \mathcal{M})$ with the quark mass matrix $\mathcal{M}$. 
From the present experiment of the electric dipole moment of the neutron, $\bar{\Theta}$ has to be smaller than $10^{-9}$. 
In SM, however, there are no reasons that $\bar{\Theta}$ should be so small, because it just a parameter of the Lagrangian. This is well known as the strong CP problem. 
One attractive solution to this problem is to introduce an anomalous global U(1) symmetry, called Peccei-Quinn (PQ) symmetry, $\rm{U(1)}_{\rm{PQ}}$ \cite{Peccei:1977hh}. When the U(1)$_{\rm{PQ}}$ is spontaneously broken, a pseudo Nambu-Goldstone boson, called axion $a$, appears. Since the axion couples with the gluon as 
\begin{eqnarray}
   \mathcal{L} = \frac{g_3^2}{64 \pi^2} 
                 \left( \bar{\Theta} + \frac{a}{f_a} \right)
                 \epsilon^{\mu \nu \rho \sigma} G^a_{\mu \nu} G^a_{\rho \sigma},
\end{eqnarray}
where $f_a$ is the decay constant of the axion, the potential of the axion is minimized at $\bar{\Theta} + \langle a \rangle/f_a = 0$. Thus, PQ symmetry can solve the strong CP problem \cite{Wilczek:1977pj}--\cite{Dine:1981rt}. 
From the above argument, it seems reasonable to consider the supersymmetric axion model, which is studied by many authors \cite{Rajagopal:1990yx}--\cite{Banks:2002sd}. 
In some supersymmetric axion models, the axino, which is a superpartner of the axion, can be  LSP and become the DM candidate \cite{Chun:1992zk}--\cite{Covi:2001nw}. 
It is realized by introducing only a Yukawa coupling  between an axion superfield and messengers 
in the superpotential \cite{Pomarol:1999ie}. In such a model, the axion superfield is a flat direction at the tree level, and hence the axino mass is induced by quantum corrections. Thus, the axino can become lighter than the neutralinos. 

In this paper, we show that the axionic mirage mediation can solve not only the $\mu$-/$B \mu$-problem but the cosmological moduli problem simultaneously. 
First, we briefly review the mirage mediation in section \ref{The mirage mediation}. 
In section \ref{The model}, we discuss about our model and show that the PQ scale can be obtained within, so-called,  the axion window by stabilizing the axion superfield. 
Throughout this paper, we will focus on the case where the axion superfield is stabilized at $10^{10}$ GeV. 
The $\mu$-problem is discussed in section \ref{mu problem}. In this section, we also find that there is no SUSY CP problem in our model. 
The cosmological implication is devoted to section \ref{cosmology} and \ref{results}. 
Since the modulus decay releases a huge amount of entropy, unwanted particles existed before  is diluted away, and thus we discuss only after the modulus decay.  
Although the modulus produces a number of gravitinos, it will cause no cosmological problem because the next LSPs (NLSP) produced by the gravitinos decay into axino LSPs. We consider the cases where NLSP is the bino, the higgsino, the stau, the stop or the wino, respectively. 
By discussing  several processes to produce the axino by non-thermally, we find that the axino with mass $\mathcal{O}(100)$ MeV produced such a way can explain the present DM abundance in any NLSP cases, if the annihilation process of NLSPs produced by the gravitino decay is not effective. 
The saxion, the real scalar component of the axion superfield, would be also harmless because the modulus decay dilutes its decay products since the lifetime of the saxion is shorter than that of the modulus due to the  relatively small decay constant $f_{\rm PQ} \simeq 10^{10}$ GeV. 
We briefly comment on whether the axion can constitute DM when the axion superfield is stabilized at $10^{12}$ GeV. Section \ref{Summary} is devoted to summary.\footnote{Introduction of a singlet field 
to solve the $\mu$-/$B\mu$-problem and the cosmological problem was also considered 
in \cite{Asaka-Yamaguchi}. The latter problem was solved by thermal inflation.}

\section{Mirage mediation of supersymmetry breaking}  \label{The mirage mediation}

First we briefly review the mirage mediation of the SUSY breaking \cite{Choi:2004sx}--\cite{Choi:2006im}.
Let us consider the following supergravity (SUGRA) $\Omega$ function and superpotential \footnote{Throughout this paper, we apply the Planck unit $M_{\Pl} \simeq 2.4 \times 10^{18} ~{\rm GeV} = 1$ unless we explicitly mention.}:
\begin{gather} \label{f function and superpotential}
   \Omega = -3 (X+X^\dagger) +(X+X^\dagger)^{q_I} |Q_I|^2,  \\
   W = W_0(X)+\frac{1}{6} \lambda_{IJK}Q_I Q_J Q_K,  
\end{gather}
where $X$ and $Q_I$ represent the modulus field and the chiral matter fields,
respectively.  
The modulus superpotential, $W_0(X)$, which
is responsible for stabilization of moduli, arises from non--perturbative
effects like gaugino condensation or stringy instanton as well as with a
modulus independent term from various origin.
In the KKLT construction \cite{Kachru:2003aw}, it is given by $W_0 = A \exp(-bX) + C$
\footnote{Here we consider the single modulus case only. Extention to the multi-modulus case is straightforward.}
. 
The function $\Omega$ has the relation to the  K\"ahler  potential, $K$, via 
\begin{eqnarray}
   \Omega = -3 e^{-K/3}.
\end{eqnarray}
From the SUGRA $\Omega$ function and the superpotential, the
 chiral/gauge fields sector of the
 SUGRA Lagrangian
 can be written by using  the compensator, $\Phi$, 
\begin{eqnarray} \label{eq: general lagrangian}
   \mathcal{L} &=& \int d^4 \theta ~ \Phi^\dagger \Phi \, \Omega 
                   + \int d \theta^2 \left[ \Phi^3 W + \frac{1}{4} 
                   f_a(X) {\mathcal{W}^a}^\alpha \mathcal{W}^a_\alpha
                   \right] + \mathrm{h.c.},  
\end{eqnarray}
where $f_a(X)$ is the gauge kinetic function.
Assuming the shift symmetry, $X \to X + i\omega$, 
in absence of $W_0(X)$, its form is restricted to $f_a(X)= p_a X + q_a$.
After integrating out $F_\Phi$, the scalar part
recovers the conventional SUGRA potential in the Einstein frame
by choosing $\langle \Phi \rangle  = \langle e^{K/6} \rangle$,
\begin{eqnarray}
V_{\rm SUGRA} &=& e^K (K^{I\bar{J}} D_I W \overline{D_J W} -3 |W|^2),
\end{eqnarray}
with $F_\Phi/\langle \Phi \rangle = K_I F^I/3 + m_{3/2}$. 
Here, a SUSY configuration, $F^I = -e^{K/2}K^{I\bar{J}} \overline{D_J W} = 0$,  
gives the extremum of $V_{\rm SUGRA}$ if it exists.
It is often the case that this corresponds to the minimum of the
potential and ends up with the AdS vacuum, $V_{\rm SUGRA} = -3 e^K |W|^2 = -3 |m_{3/2}|^2$.
To achieve observed approximately Minkowski vacuum with positive cosmological
constant, we introduce the uplifting potential,
 $V =V_{\rm SUGRA}+V_{\rm lift} \approx 0^+$. This additional SUSY breaking potential
 could be remnant of stringy  effects like the anti-D3
 brane in  the original KKLT construction  \cite{Kachru:2003aw}
 or field theoretic ``hidden sector''
 effects separated from the first term of $V_{\rm SUGRA}$ (F-term uplifting) \cite{Lebedev:2006qq}--\cite{Abe:2006xp}.
Including such a modification,  the   initial   SUSY configuration
of the modulus field, $D_X W_0(X) = 0$, shifts by $|\delta X| \sim
M_{Pl}\, (m_{3/2}^2 /m_X^2)$ and this
induces modulus SUSY breaking, $|F_X/2X_R| \simeq m_{3/2}^2/m_X$ where
 $X_R= {\rm Re}(X)$.
It is known that the non--perturbative stabilization  predicts
significant enhancement of $m_X/m_{3/2}$ due to derivative of the
exponential factors in $W \sim m_{3/2}$.  Thus $F_X/2X_R$ is
considerably suppressed relative to $m_{3/2}$. This means that, in the visible sector, 
 the anomaly mediated SUSY
 breaking of order $(F_\Phi/\langle \Phi \rangle)/8\pi^2 \approx m_{3/2}/8\pi^2$
 is equally important as the modulus mediated contribution.
The relative phase of two sources of the SUSY breaking potentially causes
 the sever SUSY CP problem \cite{Endo:2003hj}.
However, if the complex phases in $W_0$ can be rotated away by
symmetries
 which are explicitly broken by $W_0$   ({\it e.g.} the R-symmetry and the shift symmetry), the relative phase in
$F_X/(F_\Phi/\langle \Phi \rangle)$ vanishes, which is the case in the KKLT set-up \cite{Choi:1993yd} \cite{Endo:2005uy} \cite{Choi:2005uz}.
This new class of the SUSY breaking is
 dubbed ``mirage mediation'' after a peculiar renormalization group  (RG) 
 behavior of its mass spectrum which we will discuss later.

Any direct coupling with
 the source of the SUSY breaking in the K\"ahler metric undermines such a loop suppressed effect in the scalar sector.
Thus the sequestering of the SM fields and the hidden sector is essential for the mirage mediation.
Actually, it is known that naive geometrical separation only does not guarantee
 the sequestering in string theory
\cite{Anisimov:2001zz} \cite{Anisimov:2002az}.
However, strongly warped geometry, which
 is the heart of the original KKLT proposal, fulfills this requirement
\cite{Kachru:2003aw} \cite{Choi:2005ge} \cite{Choi:2006bh} \cite{Brummer:2006dg} \cite{Kachru:2007xp}.  

In literature the following  constant is defined to parametrize
 the relative strength of   the two SUSY breaking  contributions,
\begin{equation}  \label{alpha parameter}
\alpha \equiv \frac{m_{3/2}}{(F_X/2X_R) \ln(M_{Pl}/m_{3/2})} \approx \frac{m_{3/2}}{4 \pi^2 (F_X/2X_R)}.
\end{equation}
The KKLT set--up  with the uplifting potential generated by the anti--D3 brane on the tip of the warped throat 
 predicts $\alpha = 1 +{\cal O}(1/4\pi^2)$, while
there are variety of proposals which lead to different values of $\alpha = {\cal O}(1)$ \cite{Choi:2004sx}--\cite{Choi:2005uz} \cite{Abe:2005rx} \cite{Abe:2006xp}.

The soft SUSY breaking terms of the visible fields are parametrized as,
\begin{equation}
{\cal L} = -\frac{1}{2} m_{\gaugino} \overline{\gaugino} \gaugino 
-m_I^2 |\tilde{\phi}_I|^2 +A_{IJK} Y_{IJK} \tilde{\phi}_I \tilde{\phi}_J \tilde{\phi}_K + {\rm h. c.},
\end{equation}
where $\gaugino$ denotes the gaugino and 
$\tilde{\phi}_I$ is the scalar component of $Q_I$  in  canonical normalization,  while 
$Y_{IJK} \equiv \lambda_{IJK}/\sqrt{(X+X^\dag)^{q_I+q_J+q_K}}$ represents the
canonical Yukawa coupling constant.
We perform the redefinition of the matter fields, $\hat{Q}_I \equiv \Phi Q_I$
to identify the effect of the anomaly mediation,
\begin{eqnarray}
     {\cal L}          &=& \int d^4 \theta 
                      \left[ 
                         -3(X+X^\dag)|\Phi|^2+
                         (X + X^\dagger)^{q_I} |\hat{Q}_I|^2  
                      \right] \nonumber \\
              && + \int d^2 \theta 
                      \left[ 
                         \Phi^3 W(X,\hat{Q_I}/\Phi) 
                         + \frac{1}{4} f_a(X) {\mathcal{W}^a}^\alpha \mathcal{W}^a_\alpha 
                      \right] + \mathrm{h.c.}
\end{eqnarray}
It is straightforward to derive the canonical gaugino mass from the  above  gauge
kinetic function,
\begin{equation}
\label{tree-gaugino}
m_{\gaugino} = F_X \partial_X \ln Re(f_a(X)) \equiv c_{\gaugino}\, \frac{F_X}{2X_R},
\end{equation}
while the other soft SUSY breaking terms are read off
 by substituting the equation of motion for the  $F$-component of the light
 fields, $\hat{Q}_I$,
into the above Lagrangian,
\begin{equation} \label{eq:modulus mediation scalar}
m_I^2 = q_I \left|\frac{F_X}{2X_R}\right|^2,~~~~~A_{IJK} = (q_I+q_J+q_K) \frac{F_X}{2X_R}.
\end{equation}
 Here we used the fact that $\lambda_{IJK}$ is modulus independent due to
 the shift symmetry. 
It is noted that this tree-level expression derived at the cut-off scale
$\Lambda$ of the SUGRA description does not have contribution from $F_\Phi$ because of the classical invariance of the Lagrangian under the super-Weyl transformation.
However, since the super-Weyl transformation is anomalous
\cite{Randall:1998uk}--\cite{Bagger:1999rd}, we have $F_\Phi$ dependent
terms at one-loop level via,
\begin{equation}
f_a(X)\to f_a\left(\frac{\mu}{\Phi},X\right), ~~~~~(X+X^\dagger)^{q_I}
\to (X+X^\dagger)^{q_I} Z_I\left(\frac{\mu}{|\Phi|}, X+X^\dagger \right),
\end{equation}
where  $\mu$ represents the renormalization scale.
This is the anomaly mediation \cite{Randall:1998uk} \cite{Giudice:1998xp} which is loop suppressed and works whenever $F_\Phi \neq 0$
 even without the modulus contribution.
Including this effect, it is noticed that the gaugino mass reproduces
 the original tree-level
value at, so-called, {\it  mirage messenger scale}, $M_{\rm mms} = \Lambda (m_{3/2}/M_{Pl})^{\alpha/2 c_{\gaugino}}$,
 which is independent of the choice of gauge group  if $c_{\lambda_a}$ is universal  \cite{Choi:2005uz}.
This is a consequence of cancellation between the quantum correction
to the modulus mediation and the anomaly mediation.
The scalar mass parameters also show similar behavior at one-loop level
 if the relevant Yukawa  couplings are  negligible or $c_{\gaugino}=1$,
 $q_I+q_J+q_K=1$ are satisfied.
The name of  the {\it mirage} mediation is taken from the fact that 
the scale, $M_{\rm mms}$, does not correspond to any physical threshold,
although the soft parameters show the systematic pattern
(eq.(\ref{tree-gaugino}), (\ref{eq:modulus mediation scalar})) at this scale,
suggesting enhanced symmetry.

\section{The model}  \label{The model}

Let us consider the hadronic axion model \cite{Kim:1979if} in the above set-up 
by introducing the axion superfield $S$ and
 $N$ pairs of messenger fields $\Psi$ and $\bar{\Psi}$, $Q_I = (S, \Psi, \bar{\Psi}, Q_i)$, where we use a small subscript, $i$, to denote  matter fields in MSSM.
Here, $S$ is completely singlet under any unbroken gauge symmetry and $\Psi$
and $\bar{\Psi}$ are vectorlike representation of the SU(5) gauge group.
We assigned the PQ charge as $Q_{\rm{PQ}}(S) = -2$, $Q_{\rm{PQ}}(\Psi) = Q_{\rm{PQ}}(\overline{\Psi}) = 1$ and $Q_{\rm{PQ}}(X) = 0$. 
The superpotential has Yukawa coupling,
\begin{equation}  \label{eq: superpotential}
W = \lambda \hat{S} \hat{\Psi} \hat{\bar{\Psi}} + \cdot \cdot \cdot,
\end{equation}
which is allowed by the PQ symmetry.
$S$ is a flat direction in SUSY limit, however, it is lifted by
the SUSY breaking,  $F_{X,\Phi}$, caused by the mirage mediation.
We assume that $S$ is stabilized far away from the origin, breaking the
PQ symmetry. This gives a huge mass for the messengers, so that they are
integrated out at $\mu = |\langle \hat{S} \rangle| \simeq f_{\rm PQ}$, 
leaving $\hat{S}$ dependence in $f$ and $Z_I$ at low energy \cite{Pomarol:1999ie},
\begin{equation}
f\left(\frac{\mu}{\Phi},\frac{\hat{S}}{\Phi},X\right),~~~~~
Z_i\left(\frac{\mu}{|\Phi|},\sqrt{\frac{\hat{S}^\dag \hat{S}}{\Phi^\dag \Phi}},X+X^\dag\right).
\end{equation}
This introduces interactions among saxion/axino and the SM fields
suppressed by $f_{\rm PQ}$, which is much stronger than those by $M_{\Pl}$.
Later we will see these interactions play important roles in cosmology.

\begin{figure}[b]  
   \begin{center}
   \includegraphics[width=15cm, clip]{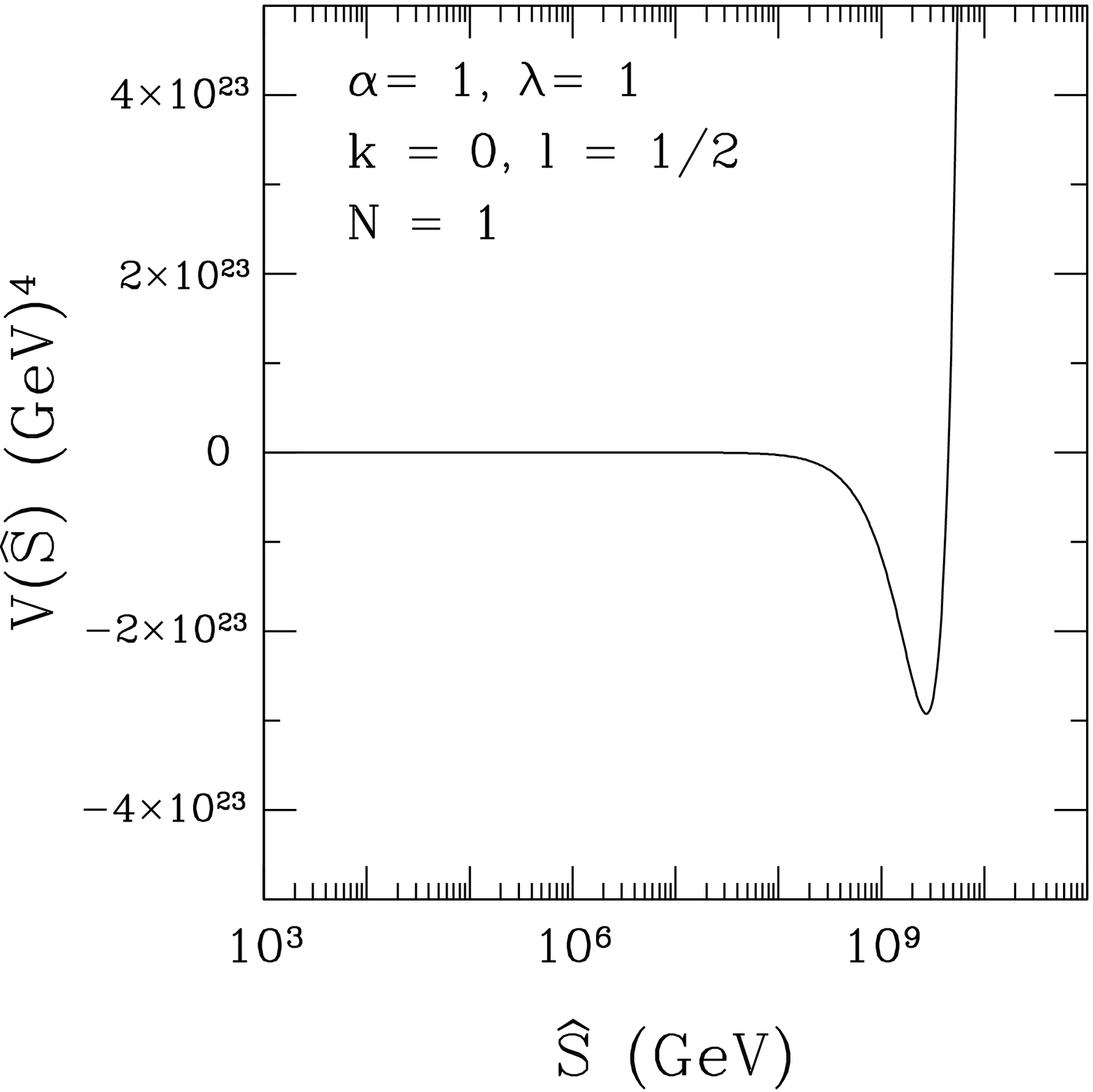} 
   \caption{The saxion potential in the axionic mirage
    mediation. The minimal model, $f\propto X$ ($c_{\gaugino} = 1$) is
    considered with $\alpha=1$, $k=0$, $\ell=1/2$, $\lambda=1$, and $N=1$, which satisfies the mirage condition.} 
\label{fig:saxion potential}
   \end{center}
\end{figure}

The scalar potential of $S$ is derived from the K\"ahler potential,
\begin{eqnarray}  \label{eq: wave function renormalization of S}
   \mathcal{L} = 
   \int d^4 \theta (X + X^\dagger)^k 
                   Z_S \left( 
                          \sqrt{ \frac{\hat{S}^\dagger \hat{S}}{\Phi^\dagger \Phi}}, \, X + X^\dagger 
                       \right) 
                   | \hat{S} |^2,
\end{eqnarray}
where $Z_S$ is the wave function renormalization of $S$ 
at $\mu = |\hat{S}|$
and  $k \equiv q_S$. We can obtain the relation among $F$-terms from the equation of motion of $F_{\hat{S}}$ as
\begin{eqnarray}  \label{eq: relation among F term}
   \frac{F_{\hat{S}}}{\hat{S}} \simeq - k\frac{F_X }{2 X_R} 
                                        + \frac{1}{2} \frac{\partial \ln Z_S}{ \partial \ln |\hat{S}|} F_\Phi.
\end{eqnarray}
Here and in what follows, we  use an abbrebiation, $F^\Phi/\langle \Phi \rangle \to F_\Phi$ 
for the sake of brevity.
It is noted that $F_X/2X_R$ is smaller than $F_\Phi$ by one-loop factor in
the ballpark, while $\partial \ln Z_S/\partial \ln|\hat{S}|=\gamma_S/(8\pi^2)$
corresponds to the anomalous dimension of $S$ at $\mu=|\hat{S}|$
 suppressed by the same factor.
Integrating  out the auxiliary component of $S$, the scalar potential is obtained as
\begin{eqnarray}  \label{eq: scalar potential of S}
   V(|\hat{S}|) \simeq  
                \left\{ k \left| \frac{F_X}{2 X_R} \right|^2 
                       - \frac{1}{4} \frac{\partial^2 \ln Z_S}{\partial (\ln |\hat{S}|)^2} \left|F_\Phi\right|^2 
                       + \frac{1}{2} \bigg( \frac{\partial^2 \ln Z_S}{\partial X \partial \ln |\hat{S}|} 
                         {F_{\Phi}}^\dagger F_X + \rm{h.c.} \bigg)
                \right\} |\hat{S}|^2 = m^2_S |\hat{S}|^2.
\end{eqnarray}
The extremum of this potential is determined by the condition,
\begin{equation}  \label{eq: extremum condition}
\frac{\partial V(|\hat{S}|)}{\partial |\hat{S}|} =\left[ \left(2 + \frac{\partial}{\partial \ln
				    |\hat{S}|}\right)m^2_S\right] |\hat{S}| = 0.
\end{equation}
If $m^2_S(\mu)$ crosses zero at some scale $\mu = \mu_0$, we expect this
condition holds in the vicinity of  $|\hat{S}|=\mu_0$ since $\partial m^2_S/\partial \ln |\hat{S}| \sim m^2_S/8\pi^2 \ll m^2_S$.
In the case of the deflected anomaly mediation \cite{Pomarol:1999ie} 
 which does not include
the modulus contribution we have introduced, the $S$ potential is given by
the second term in the parenthesis of eq.(\ref{eq: scalar potential of S})
 \cite{Abe:2001cg},
\begin{eqnarray}  \label{anomalous dimension}
   - \frac{1}{4} \frac{\partial^2 \ln Z_S}{\partial (\ln |\hat{S}|)^2} \left|F_\Phi\right|^2
      \simeq - \left( \frac{1}{16 \pi^2} \right)^2 N 
               \left[ 16 g^2_3(S) - 5 (5N + 2) \lambda^2(S) \right] \lambda^2(S) \, \left|F_\Phi\right|^2, 
\end{eqnarray}
where we have neglected the gauge coupling constants except the strong
coupling, $g_3$, and $\lambda$ is normalized canonically as $\lambda^2(S) = \lambda^2/(X + X^\dag)^{k  +q_\Psi + q_{\bar{\Psi}}} Z_S Z_{\Psi} Z_{\bar{\Psi}}$. 
It is known that this form does not change by the  RG  evolution
\footnote{Precise treatment requires a separation of the triplet and 
doublet  components in $\Psi$/$\hat{\Psi}$ and $\lambda$. We take this
into account in the following numerical  calculation.}. 
At a large value of $S$, the $\lambda^4$ term dominates
and it leads to $m^2_S > 0$, while at a small value, $m^2_S$ seems to
become negative because of the asymptotic freedom of $g_3$.
Therefore at first sight $S$ is expected to be stabilized at some scale
$\langle S \rangle$ between the cut-off and the electroweak scale.
However, in the case of the deflected anomaly mediation, 
we needed more
than 3 flavors of messengers in order to avoid the tachyonic slepton
\cite{Pomarol:1999ie}.  
Since the introduction of many number of messengers spoils
the property of the asymptotic freedom, such a stabilization mechanism
will not work without introducing new gauge interactions other than  the  SM.
 On the other hand, in our case, 
 it is enough to introduce only one or two messengers because the
 tachyonic slepton problem has already been solved by the modulus
 mediation contribution.
Furthermore, the positive contributions to the
 messenger masses, $m^2_{\Psi,\bar{\Psi}}$ and $A$-term, $A_{S\Psi\bar{\Psi}}$ from  the  modulus mediation (eq.(\ref{eq:modulus mediation scalar})) drive $m^2_S$ to negative more strongly
 than those of the deflected anomaly mediation as  read from  the RG equation,
\begin{equation}  \label{eq: RGE of m_S squared}
\frac{d m^2_S}{d \ln \mu} = \frac{5 N \lambda^2}{8\pi^2}
\left(m^2_\Psi + m^2_{\bar{\Psi}}+m^2_S + |A_{S\Psi\bar{\Psi}}|^2 \right).
\end{equation}
 Thus, the potential eq.\eqref{eq: scalar potential of S} may have a
 minimum at a scale $\langle S \rangle$. 
Actually, if $c_{\gaugino} =1, k=0$ and  $q_\Psi+q_{\bar{\Psi}}=1$ are
 satisfied, $m_S^2$ crosses zero at $M_{\rm mms}$  irrelevant to the other parameters
 in the model as discussed in the previous section \cite{Choi:2005uz}.
According to the RG equation this is at least a  local minimum as far as
 $m_\Psi^2+m_{\bar{\Psi}}^2 \gtrsim 0$, which is likely satisfied with 
the enhancement by  the strong interaction. If we increase (or decrease)
 $q_\Psi+q_{\bar{\Psi}}$ breaking the mirage condition,
 the minimum will sit above (or below) $M_{\rm mms}$.
In fig.\ref{fig:saxion potential} we show the $S$ potential for the gauge
 kinetic function, $f\propto X$ ($c_{\gaugino}=1$) with $N=1$, $k=0$,
 $q_\Psi = q_{\bar{\Psi}} \equiv \ell = 1/2$ and $\alpha=1$. The minimum comes slightly below $M_{\rm mms}$ due to the one loop correction in the derivative of the $S$ potential.

\begin{figure}[b]  
   \begin{center}
   \includegraphics[width=15cm, clip]{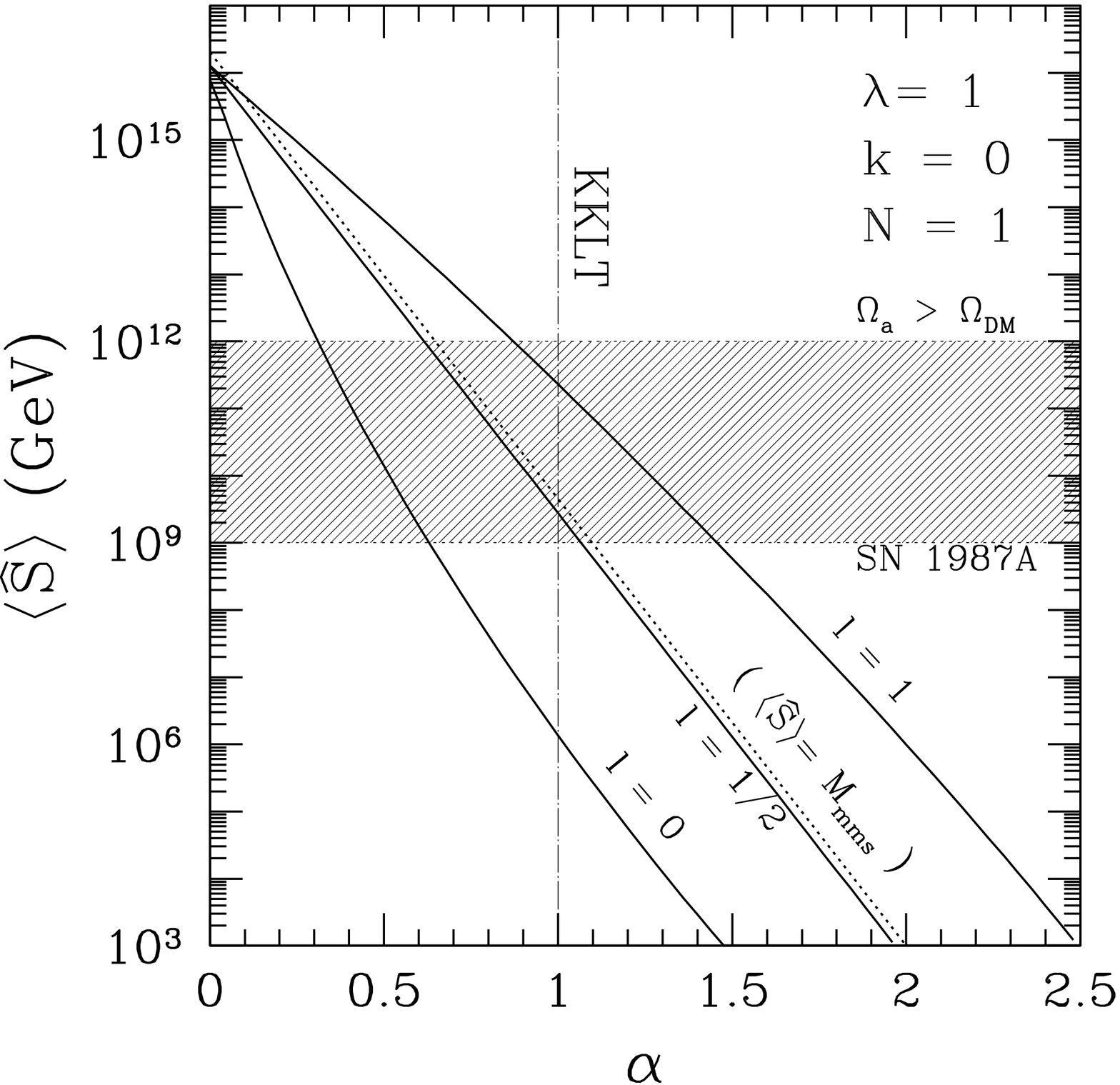} 
   \caption{The PQ scale $f_{\rm PQ}\simeq \langle \hat{S} \rangle $ in the
  axionic  mirage mediation.
The minimal model, $f_a\propto X$ ($c_{\gaugino} = 1$) is assumed.
Three different cases, $\ell=0, 1/2, 1$ are shown as a
    function of $\alpha$ for $k=0$.
$M_{mms}$ is the mirage messenger scale at which $m^2_S(\ell=1/2)$  
 crosses zero.
The remaining parameters are fixed to $\lambda=1$ and
    $N=1$, for which $f_{\rm PQ}$ shows quite modest dependence. 
For $k>0$, $f_{\rm PQ}$ is considerably lowered
 than depicted in small $\lambda$ region ($\lesssim 1$) due to the modulous contribution to $m^2_S$.
} 
\label{fig:PQ scale}
   \end{center}
\end{figure}

The PQ scale $f_{\rm{PQ}} \simeq \langle S \rangle$ is constrained by various astrophysical and cosmological considerations. 
 The relic abundance of the axion is given as \cite{Yao;2006} 
\begin{eqnarray}  \label{axion abundance}
   \Omega_a h^2 \simeq 0.7 \left( \frac{f_{\rm{PQ}}}{10^{12} \, \rm{GeV}} \right)^{7/6} 
                       \left( \frac{\Theta}{\pi} \right)^2,
\end{eqnarray}
where $\Theta$ is the misalignment angle and $h$ the Hubble constant in unit of 100 Km/sec/Mpc. 
Apart from our ignorance of $\Theta$, to avoid the axion overclosure, $f_{\rm{PQ}} \lesssim 10^{12 - 13} \, \rm{GeV}$. On the other hand, the cooling of the SN 1987A puts a lower bound, $f_{\rm{PQ}} \gtrsim 10^{9} \, \rm{GeV}$ \cite{Janka:1995ir}. 
From these observations, $f_{\rm{PQ}}$ should be in the range of $10^{9}
\, {\rm{GeV}} \lesssim f_{\rm{PQ}} \lesssim 10^{12 - 13} \,
{\rm{GeV}}$.
Let us examine whether this bound can be satisfied by our model or not.
In fig.\ref{fig:PQ scale}, we depict $\langle S \rangle$ as a function of $\alpha$ for
$c_{\gaugino}=1$, $N=1$, $k=0$ and three rational choices of $\ell = 0, 1/2, 1$.
In the KKLT set-up with the uplifting by the anti-D3 brane 
($\alpha\simeq1$), the SN bound on $f_{\rm PQ}$ is marginally satisfied with $\ell=1/2$
because $m^2_S$ crosses zero at $M_{\rm mms} \simeq 10^{10}$ GeV
 for $\alpha\simeq 1$. While
stepping up to $\ell=1$ the observed DM could be
saturated by axion taking into account involved ambiguities.
It is plausible that $\ell$ sits in the range
$0<\ell<1$ in the string effective theory because the real part of the
modulus represents the volume of the extra--dimension where the gauge fields
propagate and $\ell$ is the scaling dimension against this volume for the matter kinetic term which couples to these gauge fields.
Combining this with the axion window we obtain a constraint, $0.3 \lesssim \alpha
\lesssim 1.5$ in this set-up. We numerically checked that
 these results are robust against changes in $N$ and $\lambda$
 because both $m^2_S(\Lambda)$
 and its RG equation depend them mostly through the overall factor
 $N \lambda^2$. 
 Once we turn on
  $k> 0$, the modulus contribution
 to $m^2_S(\Lambda)$ becomes competitive and defers its crossing
 zero in RG evolution, leading to lower $f_{\rm PQ}$. 
Thus $k=0$ is only allowed choice for $\alpha\simeq 1$ unless $\lambda$ is sufficently large ($\gtrsim 1$) to dilute the modulus contribution. 
To push up $f_{\rm PQ}$ for $k>0$ or $\alpha \gtrsim 1.5$ ($e. g.$ \cite{Pierce:2006cf} \cite{Choi:2005hd}--\cite{Choi:2006xb}),
 we may need to introduce new gauge interactions which couples to the
 messengers and enhance $m_{\Psi/\bar{\Psi}}^2$ significantly via RG evolution.

Next we estimate the physical masses of the component fields 
 in $\hat{S}$.
The scalar component is decomposed into the saxion $\sigma$ and the axion $a$, so that  $\hat{S}= f_{\rm PQ}\exp[(\sigma+ia)/\sqrt{2}f_{\rm PQ}]$.
The mass of the saxion $\sigma$ at the minimum of the potential is
given by,
\begin{eqnarray}  \label{saxion mass}
m^2_\sigma &=& \frac{1}{2}\left.\frac{\partial^2 V(|\hat{S}|)}{\partial
			   |\hat{S}|^2}\right|_{\hat{S}=\langle \hat{S} \rangle}
= \frac{1}{2}\left(1+\frac{\partial}{\partial \ln|\hat{S}|}\right)\left(2+\frac{\partial}{\partial \ln|\hat{S}|}\right)m^2_S \nonumber\\
 &\simeq&  \frac{\partial m^2_S}{\partial \ln |\hat{S}|}
\simeq \frac{5 N \lambda^2}{8\pi^2}\left(m^2_\Psi+m^2_{\bar{\Psi}}+|A_\lambda|^2 \right),
\end{eqnarray}
where we used eq.(\ref{eq: extremum condition}) to derive the second line.
Assuming that $m_\Psi^2$ and $m_{\bar{\Psi}}^2$ are in the same order of magnitude with other superparticles, $m^2_\sigma$ is one-loop suppressed against the other soft breaking mass squared.
From eq.\eqref{eq: wave function renormalization of S}, the mass of the axino $\widetilde{a}$, which is the superpartner of the axion, is calculated by
\begin{eqnarray}  \label{eq: singlino mass term}
      \mathcal{L} &=&
      \int d^4 \theta (X + X^\dagger)^k 
                   Z_S \left( 
                          \sqrt{ \frac{\hat{S}^\dagger \hat{S}}{\Phi^\dagger \Phi}}, \, X + X^\dagger 
                       \right) 
                   | \hat{S} |^2 \nonumber \\
      &=& \int d^4 \theta (X + X^\dagger)^k 
          \Bigg\{ 
             Z_S + \frac{1}{2} \frac{\partial Z_S}{\partial \ln |\hat{S}|} 
             \ln \frac{\hat{S}^\dagger \hat{S}}{\Phi^\dagger \Phi} 
         + \frac{1}{8} \frac{\partial^2 Z_S}{\partial (\ln |\hat{S}|)^2} 
             \bigg( \ln \frac{\hat{S}^\dagger \hat{S}}{\Phi^\dagger \Phi} \bigg)^2  
           + \bigg( \frac{\partial Z_S}{\partial X} F_X \theta^2 + \rm{h.c.} \bigg) 
          \Bigg\} |\hat{S}|^2 \nonumber \\ 
      &\simeq& \left[ 
                  \frac{1}{2} 
                  \frac{\partial^2 \ln Z_S}{\partial \ln |\hat{S}| \partial X^\dagger} 
                  {F_X}^\dagger 
                  \bigg\langle \frac{\hat{S}^\dagger}{\hat{S}}  \bigg\rangle 
                - \frac{1}{4} \frac{\partial^2 \ln Z_S}{\partial (\ln |\hat{S}|)^2} {F_\Phi}^\dagger 
                  \bigg\langle \frac{\hat{S}^\dagger}{\hat{S}}  \bigg\rangle
               \right] \widetilde{a} ~\widetilde{a},
\end{eqnarray}
where in the last line we have normalized the fields canonically. Thus, the axino mass is 
\begin{eqnarray}   \label{eq: mass of singlino}
m_{\widetilde{a}} &=& \frac{1}{8\pi^2} 
                      \bigg( 
                         \frac{1}{2} \frac{\partial \gamma_S(|\hat{S}|)}{\partial X^\dag} F_X^\dag 
                       - \frac{1}{4}\dot{\gamma_S}(|\hat{S}|)F_\Phi^\dag 
                      \bigg) \bigg\langle \frac{\hat{S}^\dag}{\hat{S}} \bigg\rangle \nonumber \\
                  &=& - \frac{5 N }{16 \pi^2} \lambda^2(S) A_{S \Psi \bar{\Psi}} (S) \,
                      \bigg\langle \frac{\hat{S}^\dagger}{\hat{S}}  \bigg\rangle, 
\end{eqnarray}
where $\dot{\gamma_S} \equiv (d \gamma_S/d \ln \mu)$ and $A_{S \Psi \bar{\Psi}}$ is the sum of the $A$ terms of the mirage and the anomaly mediation.  
One can find that the axino mass  arises  at two-loop order in contrast with the masses of the other fields being at one-loop. Thus it will be LSP and candidate for DM. 
We discuss later how heavy is the axino which can explain the present DM abundance.

\section{The $\mu$-/$B \mu$-problem }  \label{mu problem}
An important property of this model is that it provides a natural solution to the $\mu$-/$B \mu$-problem   (Here, we used the same notation $\mu$ for both the higgsino mass and the renormalization scale). To obtain the phenomenologically viable model, the $\mu$-term and the $B  \mu$-term should be generated at the soft mass order.   
Let us consider the following superpotential and $\Omega$ function (See, for instance,  \cite{Abe:2001cg} \cite{Pomarol:1999ie})
\begin{gather} 
   W = y_1 T H_1 H_2 + y_2 S_1 S_2 T, \label{mu-term superpotential} \\
   \Omega = |S_1|^2 + |S_2|^2 + |T|^2 + \kappa S_1^\dagger S_2 + \rm{h.c.} , \label{mu-term f function}
\end{gather}
where we introduced new singlets $S_1$, $S_2$ and $T$, whose PQ charge is assigned as $Q_{\rm{PQ}}(S_1) = Q_{\rm{PQ}}(S_2) = -2$ and $Q_{\rm{PQ}}(T) = +4$, and $y_1$, $y_2$ and $\kappa$ are constants
\footnote{ Here, one can interpret the new singlet $S_1$ or $S_2$ as the previous one $S$.}. 
For simplicity, the modular weight of singlets is set to be zero.
The assignment of PQ charge to new singlets leads to $Q_{\rm PQ}(H_1) = Q_{\rm PQ}(H_2) = -2$. 
This implies that we cannot obtain the $\mu$-term from the following K$\A$hler potential: 
$\Delta K = \xi ( X + X^\dagger)^{-n} H_1 H_2$ with constant $\xi$ \cite{Giudice:1988yz}.  
The last two terms in eq.\eqref{mu-term f function} are, in general, cannot be forbidden by the PQ symmetry 
\footnote{When we write down in the superpotential all terms that are allowed by the PQ symmetry  with minimal K$\A$hler potential, we find that those terms can be translated into the form of eq.\eqref{mu-term superpotential} and eq.\eqref{mu-term f function}. We also find that the coefficient $\kappa$ does not necessarily become small when $y_1$ and $y_2$ come to be small. }. 
As was discussed in the previous section, the Yukawa coupling of the messengers with $S_1$ yields a non-vanishing VEV for $S_1$. Then, $S_2$ and $T$ become massive and we can  integrate out $T$. Its equation of motion leads to
\begin{eqnarray}  \label{EOM of T}
   S_2 \simeq - \frac{y_1}{y_2} \frac{H_1 H_2}{S_1}.
\end{eqnarray}
Substituting eq.\eqref{EOM of T} into eq.\eqref{mu-term f function}, we can obtain 
\begin{eqnarray} \label{eq: mu-term}
   \mathcal{L} = - \kappa \, \frac{y_1}{y_2}  
                 \int d^4 \theta \, \frac{\hat{S}^\dagger_1}{\hat{S_1}} H_1 H_2 + \rm{h.c.}
\end{eqnarray}
Thus, the operator eq.\eqref{eq: mu-term} generates the $\mu$-/$B \mu$-term in canonical normaization: 
\begin{gather}
   \mu =   - \frac{\kappa}{(X+X^\dag)^{(q_{H_1}+q_{H_2})/2}} \, \frac{y_1}{y_2} \frac{F_{\hat{S}_1}^\dagger}{\hat{S}_1}
\label{mu} 
         \\
   B \mu =   \frac{\kappa}{(X+X^\dag)^{(q_{H_1}+q_{H_2})/2}} \, \frac{y_1}{y_2}  \frac{F_{\hat{S}_1}^\dag}{\hat{S}_1}\left[\frac{F_{\hat{S}_1}}{\hat{S}_1}+(q_{H_1}+q_{H_2})\frac{F_X}{2X_R}\right], 
\end{gather}
where we neglected field dependence in $\kappa$ and sub--leading terms from $Z_{H_1,H_2}$
\footnote{
This corresponds to the radiative correction to $\mu$ and $B\mu$.
Note that if $y_{1,2}$ is sufficiently small the above results correctly give the input of 
(virtual) RG running at the unification scale in MSSM,
 although these terms are actually generated at $\mu = \langle S_1 \rangle$.}
.
Since they are the same order of soft masses, the $\mu$-/$B \mu$-problem can be solved. 
It is noted that in our model there is also no SUSY CP problem. 
Since the phases of two SUSY-breaking $F$-terms, ${F_\Phi}$ and ${F_X}$, can be aligned \cite{Choi:2005ge} \cite{Endo:2005uy} \cite{Choi:1993yd},    
we find that the phase of $B$, which is obtained as
\begin{eqnarray}
   B = - \frac{F_{\hat{S}_1}}{\hat{S}_1} 
       -(q_{H_1}+q_{H_2})\frac{F_X}{2X_R}
\simeq  (k-q_{H_1}-q_{H_2})\frac{F_X}{2 X_R} 
                                      - \frac{1}{2} \frac{\partial \ln Z_S}{ \partial \ln |\hat{S}|} {F_\Phi},
\end{eqnarray}
can be rotated away simultaneously with that of the gaugino mass and $A$-term. 
Therefore, we constructed a model in which there are no $\mu$-problem as well as SUSY CP problem, and it leads to the axino LSP which is lighter than the neutralino. The appearance of the lighter LSP than the neutralino is favorable for the cosmological point of view. As we mentioned earlier, the moduli problem may be solved if there is the LSP which is lighter than the neutralino.

So far, we have assumed that the singlet is stabilized by messengers as explained in the previous section. 
However, now the extra singlets can play the role of messengers, and the light quarks have PQ charges \cite{Dine:1981rt}. 
Thus, with sufficiently large Yukawa couplings, we can also construct a model without messengers.

\section{Cosmology}  \label{cosmology}

In this section, we discuss the cosmological implications of our model, in particular we shall closely investigate how it solves the LSP over-abundance problem due to gravitino decays. 
Since contributions to SUSY breaking from $S$ and $X$ are suppressed, we need to introduce an additional source of SUSY breaking, in order to cancel the cosmological constant.  
In KKLT, SUSY is broken by an anti-${D3}$ brane.  Otherwise an additional field 
$Z$ is introduced.  If supersymmetry breaking is of dynamical origin, its mass will be much higher than the electroweak scale. In what follows, we assume this is the case. 
During the inflationary epoch, the scalar fields such as $Z$, $X$ and $\sigma$ are deviated from their true minimum with some magnitudes. After inflation, each of fields will start a coherent oscillation when the Hubble parameter, $H$, becomes comparable to its mass except for the saxion. 
The behavior of the saxion is exceptional, which we will discuss shortly. 
The oscillation energy behaves like a matter, whose energy density is red shifted more slowly than that of radiation, so that it may dominate the energy of the Universe if the initial amplitude is not so small compared with the Planck scale. 
It is plausible that the modulus has the initial amplitude of the order of the Planck scale, while that of $Z$ is not the order of the Planck scale but the intermediate scale much smaller. 
Through out this paper, we assume that $Z$ decays sufficiently fast not to dominate the Universe. 
The coherent oscillation of the modulus would commence  before the reheating due to the inflaton, $\phi_{I}$, decay is completed, if the reheating temperature is not so high. 
On the other hand, the time when the saxion starts to oscillate depends on whether the saxion is trapped at the origin or deviates from the origin during the inflation.

First, we consider the case where the saxion field is trapped at the origin during the inflation. This happens if the Hubble induced mass squared for the saxion at the  inflationary era is positive. After the
inflation, the Universe is reheated. The thermal effect through the messenger fields generates the temperature dependent effective potential for the saxion field. The resulting positive mass squared proportional to the temperature squared confines it at the origin until the temperature $T$ goes down to $T=T_C \sim 1$ TeV where the thermal effect ceases to dominate over the zero-temperature contribution eq.\eqref{saxion mass}. 
The ratio of the energy density of the saxion to the entropy density at this time is 
\begin{eqnarray}  \label{energy density of saxion}
   \frac{\rho_\sigma}{s} &\simeq& \frac{45}{4 \pi^2} 
                                  \frac{m^2_\sigma \langle \hat{S} \rangle^2}{g_* T^3_C} \nonumber \\
                         &\simeq& 1.1 \times 10^{13} ~ {\rm GeV} 
                                  \left( \frac{g_*(T_C)}{100} \right)^{-1} 
                                  \left( \frac{T_C}{1 ~ \rm TeV} \right)^{-3} 
                                  \left( \frac{m_\sigma}{100 ~ {\rm GeV}} \right)^2 
                                  \bigg( \frac{\langle \hat{S} \rangle}{10^{10} ~ \rm GeV} \bigg)^2.    
\end{eqnarray}
On the other hand, since the modulus commences the coherent oscillation at the time when $H \simeq m_X$, 
\begin{eqnarray}  \label{energy density of modulus}
   \frac{\rho_X}{s} &\simeq& \frac{45}{4 \pi^2} 
                             \frac{m^2_X M_{\Pl}^2}{g_* (m_X M_{\Pl})^{3/2}} \nonumber \\
                    &\simeq& 1.7 \times 10^{10} ~ {\rm GeV} 
                             \left( \frac{g_*}{100} \right)^{-1} 
                             \left( \frac{m_X}{10^6 ~ {\rm GeV}} \right)^{1/2}.
\end{eqnarray}
The ratios remain constant until these fields decay unless other entropy productions occur. 
Comparing eq.\eqref{energy density of saxion} with eq.\eqref{energy density of modulus}, we find 
the energy density of the saxion is much larger than that of the modulus, and hence the saxion dominates the Universe. 
The saxion mainly decays into a pair of axions with the decay width 
\begin{eqnarray}  \label{saxion's decay width}
   \Gamma_\sigma \simeq \Gamma(\sigma \to a a) 
      = \frac{1}{64 \pi} \frac{m_\sigma^3}{\langle \hat{S} \rangle^2}.  
\end{eqnarray}
Since the axions produced by the saxion behave like neutrinos, they increase the Hubble expansion rate, 
and hence, the abundance of $^4$He. 
Thus, once the saxion dominates the Universe it would upset BBN unless there is an extra entropy production ($e. g.$ thermal inflation \cite{Lyth:1995ka}) after its decay.

Next, we consider the case where the saxion is displaced from the origin during the inflation. 
In this case, the saxion has at most the initial amplitude of the order of $M_{\Pl}$, $S_{\rm in} \lesssim M_{\Pl}$,  and will start to oscillate when $H \simeq m_\sigma$. 
We should note that the thermal effect generates the effective potential with positive curvature around the origin. The effect becomes irrelevant for the field value of saxion larger than the temperature because the messenger fields acquires masses larger than the temperature and the contribution to the effective potential is suppressed. How the saxion zero-mode moves under this thermal potential would be complicated, investigation of which is beyond the scope of this paper.  Being aware of the possibility that the saxion field may be trapped at the origin, we assume that this is not the case and the saxion field follows simple damped-coherent oscillation with the initial amplitude of the order of the Planck scale.  
After the reheating of $\phi_{I}$, since the oscillation energies of $X$ and $\sigma$ soon dominate the energy density of the Universe because of its relatively long lifetime. 
Their energy densities after the reheating are 
\begin{gather}
   \frac{\rho_X(t^I_d)}{s} \simeq \frac{1}{8} T_R^I X_{\rm in}^2, \quad 
   \frac{\rho_\sigma(t^I_d)}{s} \simeq \frac{1}{8} T_R^I S_{\rm in}^2,
\end{gather}
where $t^I_d$ is the time when the inflaton decays and $T_R^I$ the reheating temperature of the inflation. 
Here we consider the case both X and $\sigma$ start coherent oscillation during the era of inflation coherent oscillation (before the reheating of the $\phi_I$).  
However, since the decay constant of the saxion is much smaller than the Planck scale, the saxion decays much 
faster than the modulus. In fact, the lifetime of the saxion and the modulus is (See, eq.\eqref{saxion's decay width} and eq.\eqref{total decay width of X})
\begin{gather}
   \tau_{\sigma} \simeq 1.3 \times 10^{-8} ~ {\rm sec.}  \label{saxion's lifetime}
                        \left( \frac{m_\sigma}{100 ~ \rm GeV} \right)^{-3} 
                        \bigg( \frac{\langle \hat{S} \rangle}{10^{10} ~ \rm GeV} \bigg)^{2}, \\
   \tau_X \simeq 6.3 \times 10^{-5} ~ {\rm sec.}  \label{modulus lifetime}
                 \left( \frac{m_X}{10^6 ~ \rm GeV} \right)^{-3},    
\end{gather}
where we have set $d_g =1$. 
Therefore, even if the saxion has initial amplitude of the order of the Planck scale, the saxion decays faster than the modulus, 
and hence the modulus dominates the Universe as long as the confinement of the saxion in the thermal potential does not occur. 
When the modulus dominates the Universe, its decay produces a large amount of entropy. 
Let us estimate how large entropy is released by the modulus decay. 
When we denote that $t^X_{\rm osc}$ the time at which the coherent oscillation of the modulus commences, the ratio of the energy density of $X$ and that of the radiation at the modulus decay time, $t^X_d$, is 
\begin{eqnarray} \label{eq: ratio of rho_X and rho_R}
   \frac{\rho_X(t^X_d)}{\rho_R(t^X_d)} 
      = \frac{\rho_X(t^I_d)}{\rho_R(t^I_d)} \frac{R(t^X_d)}{R(t^I_d)} 
      = \frac{\rho_X(t^X_{\rm osc})}{\rho_{\phi_I}(t^X_{\rm osc})} 
        \left( \frac{T^I_R}{T^X_d} \right)^{4/3}
      \simeq \frac{1}{6} X^2_{\rm in} \left( \frac{T^I_R}{T^X_d} \right)^{4/3},
\end{eqnarray}
where $T^X_d$ is the decay temperature of $X$ and $R$ the scale factor. 
Therefore, unless the initial amplitude of $X$ is not much smaller than the Planck scale the entropy increase factor $\Delta$ is 
\begin{eqnarray}
   \Delta \equiv \frac{s_{\rm after}(t^X_d)}{s_{\rm before}(t^X_d)} 
               = \left( \frac{\rho_X(t^X_d)}{\rho_R(t^X_d)} \right)^{3/4} 
               \simeq X_{\rm in}^{3/2} \frac{T^I_R}{T^X_d} \sim 10^7,
\end{eqnarray}
when, for instance, $T^I_R \simeq 10^{6}$ GeV and $T^X_d \simeq 0.1$ GeV
\footnote{The modulus decay temperature with $m_X \simeq 10^6$ GeV is of the order of 0.1 GeV. See below. }. 
This relatively large entropy increase can sufficiently erase unwanted particles produced before the modulus decay. 
The axions produced by the saxion decay are also diluted. 
The ratio of the energy density of the axion, $\rho_a$, to that of the radiation at the modulus decay can be estimated as 
\begin{eqnarray}   \label{energy densty ratio of a to X}
   \frac{\rho_a (t^X_d)}{\rho_R (t^X_d)} 
      = \frac{\rho_a (t^\sigma_d)}{\rho_X (t^\sigma_d)} \frac{R (t^\sigma_d)}{R (t^X_d)}
      \simeq \left( \frac{\tau_\sigma}{\tau_X} \right)^{2/3} 
      \simeq 3 \times 10^{-3}, 
\end{eqnarray}
where $t^\sigma_d$ is the decay time of the saxion and we have used eq.\eqref{saxion's lifetime} and eq.\eqref{modulus lifetime} in the last step. 
In order that the produced axions do not change the Hubble expansion rate, its energy density should be less than that of the one neutrino species: 
\begin{eqnarray}   \label{condition for axion}
   \frac{\rho_a}{\rho_R} \bigg|_{1 \rm MeV} \lesssim \frac{7}{43}.
\end{eqnarray}
Comparing eq.\eqref{energy densty ratio of a to X} and eq.\eqref{condition for axion}, 
the axions produced by the saxion would not spoil BBN. 
Therefore, we find that it is sufficient to discuss the cosmic evolution only after the modulus decay in the case where both the modulus and the saxion have the initial amplitude of the order of the Planck scale. In what follows, we address the implications of the modulus decay to cosmology.

The modulus mainly decays into gauge multiplets through the gauge kinetic function. (See, the second term in the last line in eq.\eqref{eq: general lagrangian}.) The decay width into gauge boson pairs and into gaugino pairs is \cite{Endo:2006zj} \cite{Nakamura:2006uc} 
\begin{eqnarray}  \label{eq: X to gg}
   \Gamma(X \to gg) = \Gamma(X \to \lambda \lambda) 
         = \frac{3}{32 \pi} \left( \frac{N_G}{12} \right) d_g^2 \frac{m_X^3}{M^2_{\Pl}},
\end{eqnarray} 
where $N_G$ is the number of the gauge bosons, which is 12 for MSSM. 
The numerical factor of order unity,  $d_g$, is defined by 
\begin{eqnarray}
   d_g 
\equiv \, \, \langle G_{X X^\dagger} \rangle^{-\frac{1}{2}} \langle f_R \rangle^{-1} 
               \bigg| \bigg\langle \frac{\partial f}{\partial X} 
                      \bigg\rangle \bigg|,
\end{eqnarray}
where the subscript $X$ represents derivative with respect to $X$ and the real part $f_R = {\rm Re} f$, respectively. 
The function $G$ is the total K$\A$hler potential defined by $G \equiv K + \ln |W|^2$.  
Remarkably it was recently recognized that the decay of $X$ into the gravitino pair is not suppressed. The decay width is given by \cite{Endo:2006zj} \cite{Nakamura:2006uc} 
\footnote{In some cases where $Z$ is light and has a minimal coupling, the mixing between $Z$ and $X$ might play an important role \cite{Dine:2006ii} \cite{Endo:2006tf}, suppressing the decay of $X$ into the gravitinos. However since $m_Z \gg m_X$, this does not work in our case.}
\begin{eqnarray} \label{eq: X to 3/2 3/2}
   \Gamma(X \to \psi_{3/2}\,  \psi_{3/2}) = \frac{d^2_{3/2}}{288 \pi} \frac{m^3_X}{M^2_{\Pl}}, 
\end{eqnarray}
where $d_{3/2}$ is also order one coefficient defined by 
\begin{eqnarray}
   \langle G_{X X^\dagger} \rangle^{-1/2} \langle e^{G/2} G_X \rangle \equiv d_{3/2} \frac{m^2_{3/2}}{m_X}.
\end{eqnarray}
The modulus also decays into the saxion pair and the axino pair. The relevant interaction for the decay into the saxion pair is  
\begin{eqnarray}  \label{eq: L_XSS}
   \mathcal{L}_{X \sigma \sigma} = - \frac{1}{2} 
                         \left[ 
                              \frac{k}{(2 X_R)^2} {F_X} \frac{\partial {F_X}^\dagger}{\partial X} 
                            + \frac{1}{2} \frac{\partial^2 \ln Z_S}
                                               {\partial X^\dagger \partial \ln |\hat{S}|} 
                              F_\Phi \frac{\partial F_X^\dagger}{\partial X}
                         \right]
                 \delta X \sigma \sigma,
\end{eqnarray}
where $\delta X \equiv X - \langle X \rangle$. In eq.\eqref{eq: L_XSS}, the first term gives the dominant contribution because $\langle \partial {F_X}^\dagger/\partial X \rangle \simeq m_X$ and $\langle {F_X} \rangle \simeq m^2_{3/2}/m_X$. The decay width can be estimated as 
\begin{eqnarray}  \label{decay width of X into saxion pair}
   \Gamma(X \to \sigma \sigma) 
       \simeq \frac{k^2}{128 \pi} \left( \frac{m_{3/2}}{m_X} \right)^4 \frac{m^3_X}{M^2_{\Pl}}.
\end{eqnarray}
The interaction of $X$ with axino comes from eq.\eqref{eq: singlino mass term}
\begin{eqnarray}   \label{eq: L_Xss}
   \mathcal{L}_{X \widetilde{a} \, \widetilde{a}} 
      \simeq \frac{1}{2} \frac{\partial^2 \ln Z_S}{\partial \ln |\hat{S}| \partial X^\dagger} \,
             \frac{\partial {F_X}^\dagger}{\partial X} 
             \delta X \widetilde{a} \, \widetilde{a}.
\end{eqnarray}
Since the derivative of the anomalous dimension with respect to $X$ is   
\begin{eqnarray}
   \frac{1}{2} \frac{\partial^2 \ln Z_S}{\partial \ln |\hat{S}| \partial X^\dagger} 
      \simeq \frac{5 N \lambda^2}{16 \pi^2},
\end{eqnarray}
the decay width into the axino pair is 
\begin{eqnarray}  \label{decay of X into an axino pair}
   \Gamma(X \to \widetilde{a} \, \widetilde{a}) \simeq   \frac{1}{128 \pi} 
                                              \left( \frac{5 N \lambda^2}{16 \pi^2} \right)^2
                                              \frac{m^3_X}{M^2_{\Pl}}.
\end{eqnarray} 
Comparing these partial decay widths, we find that the modulus dominantly decays into gauge boson/gaugino pairs with the total decay width 
\begin{eqnarray} \label{total decay width of X}
   \Gamma_X \simeq \frac{3}{16 \pi} \left( \frac{N_G}{12} \right) d_g^2 \frac{m_X^3}{M^2_{\Pl}}.
\end{eqnarray}
Thus, the reheating temperature of $X$ is 
\begin{eqnarray}  \label{eq: reheating temperature of X}
 T_R^X &=& \left( \frac{90}{\pi^2 g_*(T^X_R)} \right)^{1/4}
         \sqrt{\Gamma_X M_{\rm Pl}}
\nonumber \\
     &=& 1.5 \times 10^2 \, {\rm MeV} \left( \frac{g_*(T^X_R)}{10} \right)^{-1/4}
              \left( \frac{N_G}{12} \right)^{1/2}
              d_{g} 
             \left( \frac{m_X}{10^6 \, \rm{GeV}} \right)^{3/2} ,
\end{eqnarray}
where $g_*(T^X_R)$ is the effective degrees of freedom of the radiation at the reheating. 
Before discussing how the present DM abundance is composed, we should comment on saxions produced by the modulus decay. The branching ratio of $X$ into the saxion pair is obtained from   eq.\eqref{decay width of X into saxion pair} and eq.\eqref{total decay width of X}
\begin{eqnarray} \label{branching of X into saxion pair}
   B^X_{\sigma} \equiv Br(X \to \sigma \sigma) = \frac{k^2}{24 d^2_g} \left( \frac{m_{3/2}}{m_X} \right)^4.
\end{eqnarray}
The yield of the saxion produced by the modulus decay is 
\begin{eqnarray}  \label{yield of saxion produced by X decay}
    Y^X_\sigma &=& \frac{3}{2} \, \frac{T^X_R}{m_X} B^X_{\sigma} \nonumber \\
               &\simeq& 9.6 \times 10^{-13} \, \frac{k^2}{d_g} 
                      \left( \frac{g_*(T_R^X)}{10} \right)^{-1/4} 
                      \left( \frac{N_G}{12} \right)^{-1/2} 
                      \left( \frac{m_X}{10^6 \, \rm{GeV}} \right)^{-7/2} 
                      \left( \frac{m_{3/2}}{10^5 \, \rm{GeV}} \right)^{4}. 
\end{eqnarray} 
However, from eq.\eqref{saxion's lifetime}, we find that the saxions decay immediately when they are produced by the modulus, and hence their decay does not spoil the success of BBN.  

\begin{figure}[b]  
   \begin{center}
   \includegraphics[width=15cm, clip]{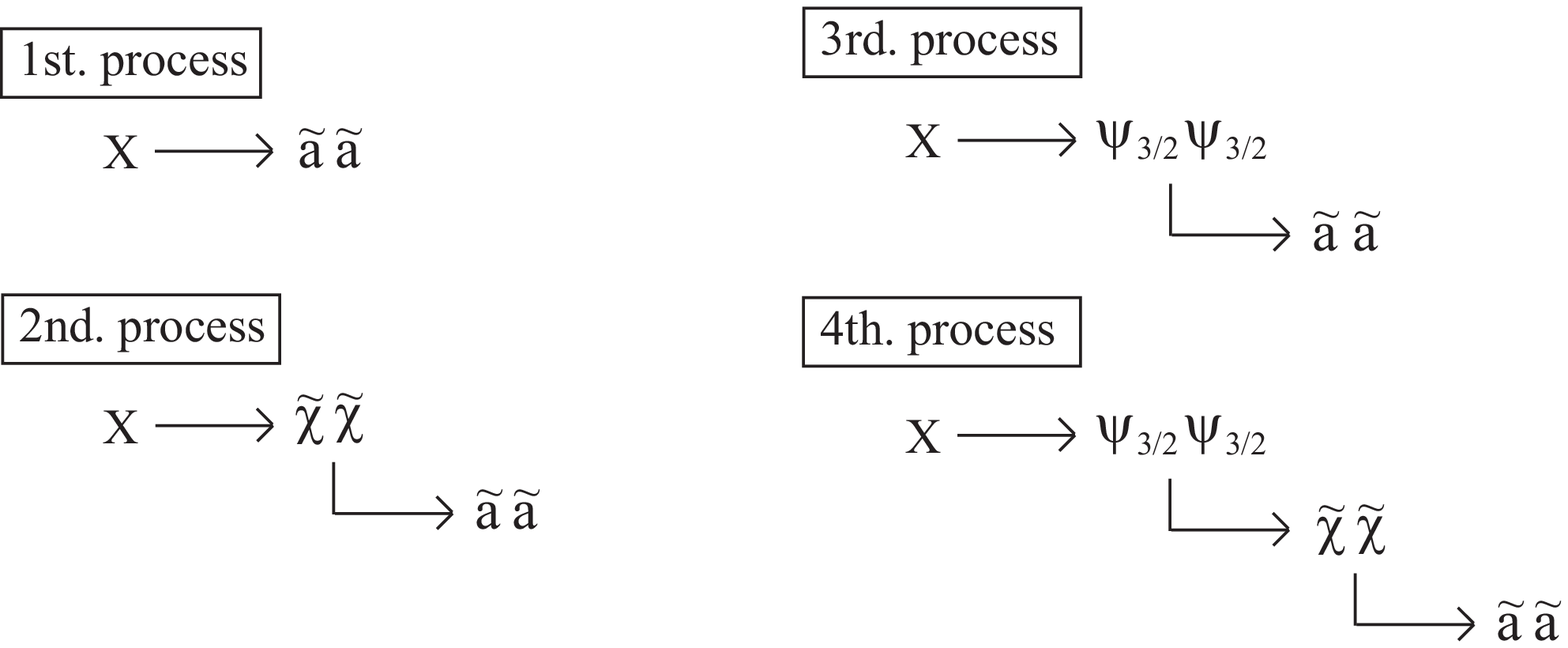} 
   \caption{The sketch of four production processes of the axino, where $\widetilde{\chi}$ denotes NLSP. The \textit{4th. process} gives the dominat contribution to the relice abundance of the axino. } 
\label{fig: figure of decay}
   \end{center}
\end{figure}

In our model, as we have already mentioned, the axino is the LSP and the DM candidate. Thus, in the following discussions, let us estimate the axino abundance and discuss whether the axino can explain the present DM abundance. 
There are four production mechanisms of axinos. We sketched these four processes in fig.\ref{fig: figure of decay}. One of the processes is the production from the decay of the modulus directly. The second comes from the decay of next LSPs (NLSP), $\widetilde{\chi}$, produced by the modulus. 
We should note that the sparticles are not in the thermal equilibrium because the reheat temperature after the X decay is not high enough. 
The abundance of axino produced by this process is essentially the same as that of NLSP, and hence we estimate the abundance of NLSP at the modulus decay.  
In the third process, axinos are produced by the decay of gravitinos. 
Since gravitinos decay uniformly into all MSSM particles and the axino, the axino abundance may be determined by the ratio of the decay channels. 
The fourth and last process is that NLSPs produced by gravitinos decay into axinos. We discuss these processes for cases where the NLSP is the bino, the higgsino and the stau in turn. 

\subsection{Bino NLSP case}
%\subsection{1st. process}
$\textit{1st. process}$. 
The branching ratio of $X$ into the axino pair can be estimated from eq.\eqref{decay of X into an axino pair} and eq.\eqref{total decay width of X} as 
\begin{eqnarray}  \label{eq: branching of X into singlino}
   B^X_{\widetilde{a}} \equiv Br(X \to \widetilde{a} \, \widetilde{a}) 
                   \simeq  \frac{1}{24 d^2_g}
                          \left(\frac{5 N \lambda^2}{16 \pi^2} \right)^2.
\end{eqnarray}
According to eq.\eqref{eq: branching of X into singlino} and eq.\eqref{eq: reheating temperature of X}, the relic abundance of axinos directly produced by the modulus decay is 
\begin{eqnarray}  \label{axino abundance comes from modulus}
   Y^X_{\widetilde{a}} = \frac{3}{2} \frac{T^X_R}{m_X} B_{\widetilde{a}}^X 
                 \simeq 9.3 \times 10^{-12} N^2 \lambda^4 \left( \frac{g_*(T^X_R)}{10} \right)^{-1/4}
                        d_g^{-1} \left( \frac{N_G}{12} \right)^{1/2}
                        \left( \frac{m_X}{10^6  \, \rm{GeV}} \right)^{1/2}.
\end{eqnarray}
For the axino mass below the electroweak scale, this yield will be too small to account for the present DM abundance.

%\subsection{2nd. process}
$\textit{2nd. process}$.
When the modulus reheats the Universe, neutralinos are so abundant that the annihilation among them may
become effective. 
However, since the neutralino can decay into the axino, if the decay width is larger than the interaction rate, then the annihilation process will not be effective.  
The bino decays into the axino and a photon or a $Z^0$-boson through the following coupling
\begin{eqnarray}  \label{coupling of bino with axino and photon}
   \mathcal{L}_{\widetilde{\chi} \widetilde{a} \gamma} 
      = \frac{\alpha_1 N}{16 \sqrt{2} \pi} \frac{1}{\langle \hat{S} \rangle} \,
        \overline{\widetilde{a}} \gamma_5 \left[ \gamma^\mu , \gamma^\nu \right] 
        \widetilde{B} B_{\mu \nu},
\end{eqnarray}
where $N$ is the number of the messengers, $B_{\mu \nu}$ the field strength of the U(1)$_Y$ gauge boson $B_\mu$. In this expression, we used the four component notation. Eq.\eqref{coupling of bino with axino and photon} leads to \cite{Covi:2001nw}
\begin{subequations}  \label{decay width of bino1}
\begin{gather}  
   \Gamma( \widetilde{B} \to \widetilde{a} \gamma) 
              \simeq \frac{\alpha^2_{\rm{em}} N^2}{256 \pi^3} \frac{1}{\cos^2 \theta_W}  
                  \frac{m^3_{\widetilde{B}}}{\langle \hat{S} \rangle^2},  \\
   \Gamma( \widetilde{B} \to \widetilde{a} Z^0) 
              \simeq \frac{\alpha^2_{\rm{em}} N^2}{256 \pi^3} \frac{\tan^2 \theta_W}{\cos^2 \theta_W} 
                     ( 1 - x_Z ) \left( 1 - \frac{x_Z}{2} - \frac{x_Z^2}{2} \right) 
                     \frac{m^3_{\widetilde{B}}}{\langle \hat{S} \rangle^2},
\end{gather}
\end{subequations}
where $\alpha_{\rm{em}}$ is the fine structure constant, $\theta_W$ the Weinberg angle and $x_Z \equiv m_{Z^0}^2/m_{\widetilde{\chi}}^2$ with $Z^0$-boson mass, $m_{Z^0}$, and the NLSP mass, $m_{\widetilde{\chi}}$. 
When the bino is sufficiently heavy, its lifetime is relatively short
\begin{eqnarray}  \label{lifetime of bino}
   \tau_{\widetilde{B}} &=& \Gamma_{\widetilde{B}}^{-1} \equiv 
                        \left( \Gamma( \widetilde{B} \to \widetilde{a} \gamma) + 
                               \Gamma( \widetilde{B} \to \widetilde{a} Z^0) 
                        \right)^{-1}    \nonumber \\
                    &\simeq& 1.9 \times 10^{-4} \, {\rm{sec.}},
\end{eqnarray}
for $m_{\widetilde{B}} = 300$ GeV and $\langle S \rangle = 10^{10}$ GeV 
\footnote{In the mirage mediation, such a relatively heavy bino is usually realized.}. 
Assuming that the bino is heavy enough to annihilate into a top quark pair through s-wave, 
the annihilation cross section of the bino, $\langle \sigma_{\rm{ann.}} v_{\rm{rel.}} \rangle_{\widetilde{B}}$, is given by \cite{Olive:1989jg}
\begin{eqnarray}  \label{annihilation cross section of bino}
   \langle \sigma_{\rm{ann.}} v_{\rm{rel.}} \rangle_{\widetilde{B}} 
      \simeq \frac{32 \pi}{27} \alpha^2_1 
             \frac{m^2_t}{( m^2_{\widetilde{t}_R} + m^2_{\widetilde{B}} - m^2_t )^2} 
             \left( 1 - \frac{m^2_t}{m^2_{\widetilde{B}}} \right)^{1/2}
\end{eqnarray}
with $m_t$, $m_{\widetilde{t}_R}$ and $m_{\widetilde{B}}$ being the masses of the top-quark, the right-handed stop and the bino, respectively. Here $\sigma_{\rm{ann.}}$ is the annihilation cross section of two binos, $v_{\rm{rel.}}$ their relative velocity and  $\langle \cdots \rangle$ represents the thermal average 
\footnote{If the bino is so light that it cannot annihilate into the top quark, since the annihilation cross section becomes proportional to their relative velocity, the cross section is more suppressed. In such a case, the relic abundance of the bino becomes more redundant, and hence in order to explain the present DM abundance, the axino mass should be smaller about one order or so than that in our case.}. 
Thus, from eq.\eqref{decay width of bino1} and eq.\eqref{annihilation cross section of bino}, we can find that the interaction rate at the modulus decay  is much larger than the decay width:
\begin{eqnarray}
   \langle \sigma_{\rm{ann.}} v_{\rm{rel.}} \rangle_{\widetilde{B}} \, n_{\widetilde{B}} 
      = \langle \sigma_{\rm{ann.}} v_{\rm{rel.}} \rangle_{\widetilde{B}} Y_{\widetilde{B}} \, s 
      \gg \Gamma_{\widetilde{B}},
\end{eqnarray}
where $n_{\widetilde{B}}$ is the number density of the bino. 
The annihilation process will terminate when the Hubble parameter becomes
comparable to the annihilation rate
\begin{eqnarray}
   \langle \sigma_{\rm{ann.}} v_{\rm{rel.}} \rangle_{\widetilde{B}} \, n_{\widetilde{B}} \simeq H(T^X_R).
\end{eqnarray}
The bino abundance after their annihilation is estimated as 
\begin{eqnarray}  \label{annihilation}
   \frac{n_{\widetilde{B}}}{s} \bigg|_{T^X_R} 
      \simeq \frac{1}{4} \left( \frac{90}{\pi^2 g_*} \right)^{1/2}
             \frac{1}{\langle \sigma_{\rm{ann.}} v_{\rm{rel.}} \rangle_{\widetilde{B}} T^X_R M_{\Pl}}.
\end{eqnarray}
After the annihilation, the bino NLSP decays into the axino, and hence the axino abundance is the same as that of bino:
\begin{eqnarray}  \label{axino abundance from bino NLSP}
   Y^{\widetilde{B}}_{\widetilde{a}} \simeq \frac{n_{\widetilde{B}}}{s} \bigg|_{T^X_R} 
                             &\simeq& 6.2 \times 10^{-19} \left( \frac{g_*(T^X_R)}{10} \right)^{-1/4} 
                         d_g^{-1} \left( \frac{N_G}{12} \right)^{-1/2}
                         \left( \frac{m_X}{10^6 \, \rm{GeV}} \right)^{-3/2} 
                         \frac{{\rm{GeV}}^{-2}}
                              {\langle \sigma_{\rm{ann.}} v_{\rm{rel.}} \rangle_{\widetilde{B}}} \nonumber \\
                             &\simeq& 9.4 \times 10^{-10} \left( \frac{g_*(T^X_R)}{10} \right)^{-1/4} 
                         d_g^{-1} \left( \frac{N_G}{12} \right)^{-1/2} 
                         \left( \frac{m_X}{10^6 \, \rm{GeV}} \right)^{-3/2},
\end{eqnarray}
where we have set $m_t = 174$ GeV, $m_{\widetilde{B}} = 300$ GeV and $m_{\widetilde{t}_R}/m_{\widetilde{B}} = 1.25$ as the reference values.

%\subsection{3rd. process}
$\textit{3rd. process}$.
A fraction of the gravitinos decays into axinos. Such an axino abundance is provided by the ratio of the decay channels. 
Since the total decay width of the gravitino with negligible final state mass is (see, for instance \cite{Moroi:1995fs})
\begin{eqnarray}
   \Gamma_{3/2} = \frac{244}{384 \pi} \frac{m^3_{3/2}}{M^2_{\Pl}},
\end{eqnarray}
the axino abundance can be estimated as 
\begin{eqnarray}  \label{axino abundance from direct gravitino decay}
   Y^{3/2}_{\widetilde{a}} &\simeq& \frac{1}{244} Y^X_{3/2} 
                       = \frac{1}{244} \, \frac{3}{2} B_{3/2} \frac{T^X_R}{m_X} \nonumber \\
                       &\simeq& 1.7 \times 10^{-11} \left( \frac{g_*(T^X_R)}{10} \right)^{-1/4}
                        d_g \left( \frac{N_G}{12} \right)^{1/2}
                        \left( \frac{m_X}{10^6  \, \rm{GeV}} \right)^{1/2},
\end{eqnarray}
where $Y^X_{3/2}$ denotes the yield of the gravitino produced by the modulus decay. 
In the last step, we used the branching ratio into the gravitino pair $B_{3/2}$ which is defined by 
\begin{eqnarray}
   B_{3/2} = \frac{1}{54} \frac{d^2_{3/2}}{d^2_g} \left( \frac{N_G}{12} \right)^{-1} \simeq 0.01.
\end{eqnarray}

%\subsection{4th. process}
$\textit{4th. process}$.
Finally, let us consider the process that the binos produced by the gravitinos decay into axinos. 
In distinction to the case of the modulus decay, the decay temperature of the gravitino is not so high, 
\begin{eqnarray}
    T_{3/2} &\simeq& \left( \frac{90}{\pi^2 g_*(T_{3/2})} \right)^{1/4}
                     \sqrt{\Gamma_{3/2} M_{\rm Pl}} \nonumber \\
            &\simeq& 9.0 \times 10^{-3}\, {\rm GeV}
                           \left( \frac{g_* (T_{3/2})}{10} \right)^{-1/4}
                           \left( \frac{m_{3/2}}{10^5 \, {\rm GeV}} \right)^{3/2},
\end{eqnarray}
and then, the annihilation process does not occur effectively.  
Therefore, the axino abundance is the same as that of the gravitinos:
\begin{eqnarray}  \label{axino abundance from direct decay of bino}
   Y^{\widetilde{B}}_{\widetilde{a}} \simeq Y^X_{3/2} 
                 \simeq 4.3 \times 10^{-9} \left( \frac{g_*(T^X_R)}{10} \right)^{-1/4}
                        d_g \left( \frac{N_G}{12} \right)^{1/2}
                        \left( \frac{m_X}{10^6  \, \rm{GeV}} \right)^{1/2}.
\end{eqnarray}
Comparing eqs.\eqref{axino abundance comes from modulus}, \eqref{axino abundance from bino NLSP}, \eqref{axino abundance from direct gravitino decay} and \eqref{axino abundance from direct decay of bino}, we find that the axino abundance is dominantly constituted from the $\textit{4th. process}$, that is, decay of the binos produced by the gravitinos, eq.\eqref{axino abundance from direct decay of bino}.

We should estimate whether the binos can become non-relativistic by scattering with thermal bath \cite{Kawasaki:1995cy} \cite{Hisano:2000dz}. 
At the gravitino decay, binos lose its energy mainly by scattering with the back-ground electron. When the bino is produced by the gravitino, it is relativistic, and then the cross section, $\sigma_{\rm scatt.}$, is estimated by \cite{Kawasaki:1995cy} 
\begin{eqnarray}  \label{scattering cross section of bino}
   \langle \sigma_{\rm scatt.} v_{\rm rel.} \rangle 
      \simeq 128 \pi \alpha^2_1 \frac{E_{\widetilde{B}}^2 T_{3/2}^2}{m_{\widetilde{e}_R}^4 m_{\widetilde{B}}^2},
\end{eqnarray}
where $E_{\widetilde{B}}$ is the energy of the bino and $m_{\widetilde{e}_R}$ the mass of the right-handed selectron. 
The energy loss rate for the relativistic NLSP, $\Gamma_{\rm scatt.}^{\rm NLSP}$, is given by 
\begin{eqnarray}   \label{energy loss rate for bino}
   \Gamma_{\rm scatt.}^{\widetilde{B}} \simeq n_e \langle \sigma_{\rm scatt.} v_{\rm rel.} \rangle 
                              \frac{\Delta E_{\widetilde{B}}}{E_{\widetilde{B}}},
\end{eqnarray}
where $n_e$ is the number density of the back-ground electron and $\Delta E_{\widetilde{B}}/E_{\widetilde{B}}$ the averaged energy loss rate of bino in one scattering, which is given by 
\begin{eqnarray}   
   \frac{\Delta E_{\widetilde{B}}}{E_{\widetilde{B}}} 
      \simeq  12 \bigg( \frac{E_{\widetilde{B}} T_{3/2}}{m_{\widetilde{B}}^2} \bigg). 
\end{eqnarray}
Comparing the energy loss rate eq.\eqref{energy loss rate for bino} with the decay width eq.\eqref{decay width of bino1}, we can find 
\begin{eqnarray}
   \frac{\Gamma^{\widetilde{B}}_{\rm scatt.}}{\Gamma_{\widetilde{B}}} \bigg|_{T_{3/2}}
      \sim 2 \times 10^3
\end{eqnarray}
for $m_{\widetilde{B}} \simeq m_{\widetilde{e}_R}$ = 300 GeV, $T_{3/2} \simeq$ 10 MeV and $E_{\widetilde{B}} \simeq 50$ TeV. 
Therefore, at the gravitino decay the binos decay into axinos after becoming non-relativistic without the annihilation process: 
\begin{eqnarray}
   \Gamma^{\widetilde{B}}_{\rm scatt.} \gg \Gamma_{\widetilde{B}} 
     \gg \langle \sigma_{\rm{ann.}} v_{\rm{rel.}} \rangle_{\widetilde{B}} \, n_{\widetilde{B}}.
\end{eqnarray}

\subsection{Higgsino NLSP case}
Let us next consider the case of the higgsino NLSP.  The results of the \textit{1st.} and \textit{3rd. processes}  remain valid, irrespective of the composition of NLSPs. 
A difference from the bino NLSP case is the annihilation cross section. It is given by \cite{Olive:1989jg}
\begin{eqnarray}  \label{cross section of higgsino}
   \langle \sigma_{\rm{ann.}} v_{\rm{rel.}} \rangle_{\widetilde{h}}
        \simeq \frac{g^4_2}{32 \pi} \frac{1}{m^2_{\widetilde{h}}} \frac{(1-x_W)^{3/2}}{(2-x_W)^2} 
             + \frac{g^4_2}{64 \pi \cos^4 \theta_W} 
               \frac{1}{m^2_{\widetilde{h}}} \frac{(1-x_Z)^{3/2}}{(2-x_Z)^2}
\end{eqnarray}
with $x_{W} \equiv m^2_{W}/m^2_{\widetilde{\chi}}$, $m_{W}$ being the $W$-boson mass, $g_2$ the SU(2) gauge coupling constant and $m_{\widetilde{h}}$ the higgsino mass. Here coannihilation is not taken into 
account.
The higgsino can decay into the axino and the Higgs boson through the $\mu$-term, eq.\eqref{eq: mu-term}, 
\begin{eqnarray}  \label{coupling of higgsino with Higgs and axino}
   \mathcal{L} = \kappa \frac{y_1}{y_2} \frac{1}{\langle \hat{S} \rangle} 
                 \left\{  
                    \bigg\langle \frac{F_{\hat{S}}^\dagger}{\hat{S}} \bigg\rangle 
                    \Big( 
                       \widetilde{a} \, \widetilde{h}^0_1 \, h^0_2 + \widetilde{a} \, \widetilde{h}^0_2 \, h^0_1 
                    \Big)     
                  + \rm{h.c.}
                 \right\}.
\end{eqnarray}
The physical neutral Higgs bosons are related to $h_1^0$ and $h_2^0$ by 
\begin{gather}  
   h_1^0 = \frac{1}{\sqrt{2}} \left( v_1 - h^0 \sin \alpha + H^0 \cos \alpha 
                                    + i A^0 \sin \beta + i G^0 \cos \beta 
                              \right) \\ 
   h_2^0 = \frac{1}{\sqrt{2}} \left( v_2 + h^0 \cos \alpha + H^0 \sin \alpha 
                                    + i A^0 \cos \beta - i G^0 \sin \beta
                              \right)
\end{gather}
with $v_1$ and $v_2$ being VEVs of the $H_1$ and $H_2$. Here $h^0$, $H^0$ and $A^0$ are physical scalars, which correspond to the CP even light Higgs, the CP even heavy Higgs and the CP odd Higgs, respectively, and $G^0$ is the Nambu-Goldstone boson.  The mixing angle $\alpha$ (which should not be confused with $\alpha$ parameter eq.\eqref{alpha parameter}) satisfies the relation 
\begin{eqnarray}
   \tan 2 \alpha = \frac{m^2_A + m^2_{Z^0}}{m^2_A - m^2_{Z^0}} \tan 2 \beta,
\end{eqnarray}
where $\tan \beta \equiv v_2/v_1$ and $m_A$ is the mass of $A^0$. 
Then, the decay width of the higgsino into the axino and the Higgs boson is 
\begin{eqnarray}  \label{decay width of the higgsino}
   \Gamma(\widetilde{h}_\pm \to \widetilde{a} h^0 ) 
       \simeq \frac{1}{32 \pi} \left( \cos \alpha \mp \sin \alpha \right)^2  
              \bigg( 1 - \frac{m^2_{h^0}}{m_{\widetilde{h}}^2} \bigg)^2 
              \left( \frac{\mu}{m_{\widetilde{h}}} \right)^2 
               \frac{m_{\widetilde{h}}^3}{\langle S \rangle^2}, 
\end{eqnarray}
where $\mu$ is the $\mu$-term, $i.e.$, eq.\eqref{mu} and $\widetilde{h}_\pm$ the mass eigenstate, which are defined as $\widetilde{h}_\pm \equiv \frac{1}{\sqrt{2}} ( \widetilde{h}^0_1 \pm \widetilde{h}^0_2)$. 
The higgsino can also decay into the axino and the $Z^0$ boson through the axino-higgsino mixing. 
According to eq.\eqref{eq: mu-term}, the axino and the higgsino are mixed each other by
\begin{eqnarray}   \label{axino--higgsino mixing}
   \mathcal{L}_{\rm mix} = - \kappa \frac{y_1}{y_2} 
                           \bigg\langle \frac{F_{\hat{S}}^\dagger}{\hat{S}^2} \bigg\rangle 
                           v_2 \, \widetilde{a} \, \widetilde{h}_1^0 
                           - \kappa \frac{y_1}{y_2} 
                           \bigg\langle \frac{F_{\hat{S}}^\dagger}{\hat{S}^2} \bigg\rangle 
                           v_1 \, \widetilde{a} \, \widetilde{h}_2^0 + {\rm h.c.}
\end{eqnarray}
This mixing can be removed by transformations  
\begin{eqnarray}   \label{mass eigenstates}
   {\widetilde{a}}^{\, \prime} \!\! &\simeq& \!\! 
                                  \widetilde{a} - \frac{\epsilon_1}{\mu} \widetilde{h}_1^0 
                                  - \frac{\epsilon_2}{\mu} \widetilde{h}_2^0, \\ 
   {\widetilde{h}_1}^{0 \, \prime} \!\! &\simeq& \!\! 
                                       \widetilde{h}_1^0 
                                       + \frac{\epsilon_1}{\mu} \, \widetilde{a}, \\
   \widetilde{h}_{2}^{0 \, \prime} \!\! &\simeq& \!\! 
                                       \widetilde{h}_2^0 
                                       + \frac{\epsilon_2}{\mu} \, \widetilde{a}, 
\end{eqnarray}
where we have neglected the higher power of $\epsilon_i/\mu \equiv - \kappa \dfrac{y_1}{y_2} \bigg\langle \dfrac{F_{\hat{S}}^\dagger}{\hat{S}^2} \bigg\rangle v_i/\mu \simeq v_i/\langle \hat{S} \rangle$. 
In the unitary gauge, the relevant term to the decay is 
\begin{eqnarray}  \label{interaction of higgsino to axino and Z}
   \mathcal{L}  = 
                 - \frac{g_2}{2 \cos \theta_W} \frac{\epsilon_1}{\mu} \,
                   \overline{\widetilde{h}_1^0} \, \bar{\sigma}^\mu \, \widetilde{a} \, Z_\mu 
                 + \frac{g_2}{2 \cos \theta_W} \frac{\epsilon_2}{\mu} \,
                   \overline{\widetilde{h}_2^0} \, \bar{\sigma}^\mu \, \widetilde{a} \, Z_\mu. 
\end{eqnarray}
Thus, the decay width of the higgsino into the axino and the $Z^0$ boson is 
\begin{eqnarray}  \label{higgsino decay into a and Z}
   \Gamma(\widetilde{h}_\pm \to \widetilde{a} Z^0) \simeq 
      \frac{1}{32 \pi} \big( \cos \beta \mp \sin \beta \big)^2  
                       \bigg( 1 - \frac{m_{Z^0}^2}{m_{\widetilde{h}}^2} \bigg)^2 
                       \bigg( 1 + 2 \, \frac{m_{Z^0}^2}{m_{\widetilde{h}}^2} \bigg)
                       \frac{m_{\widetilde{h}}^3}{\langle \hat{S} \rangle^2}. 
\end{eqnarray}
The lifetime of the higgsino can be estimated as 
\begin{eqnarray}
   \tau_{\widetilde{h}} = \Gamma_{\widetilde{h}}^{-1} 
      \equiv \left( 
                 \Gamma(\widetilde{h} \to \widetilde{a} h^0) + \Gamma(\widetilde{h} \to \widetilde{a} Z^0)
             \right)^{-1}
      \simeq 2.4 \times 10^{-9} \, \rm{sec.},
\end{eqnarray}
for $m_{h^0} = 120$ GeV, $m_{Z^0} = 91$ GeV, $m_{\widetilde{h}} = 150$ GeV and $\langle S \rangle = 10^{10}$ GeV.  
According to eq.\eqref{cross section of higgsino} and eq.\eqref{decay width of the higgsino} and eq.\eqref{higgsino decay into a and Z}, we can find that in the $\textit{2nd. process}$, the  annihilation is more effective than the decay. 
Since the higgsinos decay into axinos after their annihilation, the axino abundance of the $\textit{2nd. process}$ is comparable to that of the  higgsinos  
\begin{eqnarray}
   Y^{\widetilde{h}}_{\widetilde{a}} 
                 \simeq \frac{n_{\widetilde{h}}}{s} \bigg|_{T^X_R} 
                 &\simeq& 6.2 \times 10^{-19} \left( \frac{g_*(T^X_R)}{10} \right)^{-1/4} 
                         d_g^{-1} \left( \frac{m_X}{10^6 \, \rm{GeV}} \right)^{-3/2} 
                         \frac{{\rm{GeV}}^{-2}}
                              {\langle \sigma_{\rm{ann.}} v_{\rm{rel.}} \rangle_{\widetilde{h}}} \nonumber \\ 
                 &\simeq& 2.1 \times 10^{-11} \left( \frac{g_*(T^X_R)}{10} \right)^{-1/4} 
                          d_g^{-1} \left( \frac{m_X}{10^6 \, \rm{GeV}} \right)^{-3/2}.
\end{eqnarray}
Here, in the last step, we have set $m_{\widetilde{h}} = 150\, \rm{GeV}$ for reference.

On the other hand, in the $\textit{4th. process}$, the annihilation does not occur effectively as in the case of the bino NLSP. 
Therefore,  the axino abundance produced by the $\textit{4th. process}$ is the same as the gravitino abundance
\begin{eqnarray}  \label{axino abundance from direct decay of higgsino}
   Y^{\widetilde{h}}_{\widetilde{a}} \simeq Y^X_{3/2} 
                 \simeq 4.3 \times 10^{-9} \left( \frac{g_*(T^X_R)}{10} \right)^{-1/4}
                        d_g \left( \frac{N_G}{12} \right)^{1/2}
                        \left( \frac{m_X}{10^6  \, \rm{GeV}} \right)^{1/2},
\end{eqnarray}
and hence we find this is the dominant contribution to the axino relic abundance. 
As the case of the bino NLSP, the higgsinos also become non-relativistic before its decay. 
The scattering rate for the higgsino is given by \cite{Hisano:2000dz}
\begin{eqnarray}  \label{higgsino's scattering rate}
   \Gamma^{\widetilde{h}}_{\rm scatt.} &\simeq& 
      8 \pi^2 \alpha_2^2 \frac{E_{\widetilde{h}} T_{3/2}^4}{m_W^4} \, 
      e^{- m_{\widetilde{h}} \Delta m_{\widetilde{\chi}}/  2 E_{\widetilde{h}} T_{3/2}} 
      \left( 
         \frac{\Delta m_{\widetilde{\chi}}}{ m_{\widetilde{h}}} 
         + 6 \frac{E_{\widetilde{h}} T_{3/2}}{m_{\widetilde{h}}^2} 
      \right)  \nonumber \\ 
      &&\times
      \left( 
         12 \frac{E_{\widetilde{h}} T_{3/2}}{m_{\widetilde{h}}^2} 
         - 2 \frac{\Delta m_{\widetilde{\chi}}}{m_{\widetilde{h}}} 
      \right)  N_F,
\end{eqnarray}
where $\Delta m_{\widetilde{\chi}} \equiv m_{\widetilde{\chi}^+} - m_{\widetilde{h}}$ is the mass difference between the chargino and the higgsino, $E_{\widetilde{h}}$ the energy of the higgsino and $N_F$ the number of the processes. 
Taking the ratio eq.\eqref{decay width of the higgsino} to eq.\eqref{higgsino's scattering rate}, we find 
\begin{eqnarray}
   \frac{\Gamma^{\widetilde{h}}_{\rm scatt.}}{\Gamma_{\widetilde{h}}} \bigg|_{T_{3/2}} \sim 
      400 N_F,
\end{eqnarray}
for $E_{\widetilde{h}} =$ 50 TeV, $m_{\widetilde{h}} =$ 150 GeV, $\Delta m_{\widetilde{\chi}} \simeq$ 5 GeV and $T_{3/2} \simeq$ 10 MeV 
\footnote{In the other NLSP cases, one would also find the scattering process of NLSP with thermal bath is more effective than its decay width. Therefore, any NLSPs become non-relativistic before its decay.}.

\subsection{Stau NLSP case}
Next, we discuss the case for the stau NLSP. The annihilation cross section of the stau is given by \cite{Asaka:2000zh}
\begin{eqnarray}  \label{stau's annihilation cross section}
   \langle \sigma_{\rm{ann.}} v_{\rm{rel.}} \rangle_{\widetilde{\tau}} 
      \simeq \frac{4 \pi \alpha_{\rm{em}}^2}{m^2_{\widetilde{\tau}}} 
             + \frac{16 \pi \alpha_{\rm{em}}^2 m^2_{\widetilde{B}}}
                    {\cos^4 \theta_W \big( m^2_{\widetilde{\tau}} + m^2_{\widetilde{B}} \big)^2},
\end{eqnarray}
where $m_{\widetilde{\tau}}$ is the stau mass. For simplicity, we do not consider the coannihilation of $\widetilde{\tau}$ with $\widetilde{B}$. 
The stau can decay into the axino through the axino-higgsino mixing eq.\eqref{axino--higgsino mixing}. 
Thus, the coupling of the stau with the axino becomes 
\begin{eqnarray}   \label{stau's coupling}
   \mathcal{L}_{\widetilde{\tau} \tau \widetilde{a}} 
       \simeq - y_\tau \frac{\epsilon_1}{\mu} \, 
              \widetilde{\tau}^*_R \, \tau \, \widetilde{a}^{\, \prime} + {\rm h.c.}
\end{eqnarray}
with $y_\tau$ the Yukawa coupling of $\tau$. Therefore, we can obtain the decay width of the stau is 
\begin{eqnarray}   \label{stau's decay width}
   \Gamma(\widetilde{\tau} \to \widetilde{a} \tau) \simeq 
       \frac{1}{32 \pi} \left( \frac{m_\tau}{m_{\widetilde{\tau}}} \right)^2 
       \frac{m_{\widetilde{\tau}}^3}{\langle \hat{S} \rangle^2}.
\end{eqnarray}
The lifetime is estimated as 
\begin{eqnarray}  \label{stau's lifetime}
   \tau_{\widetilde{\tau}} \simeq 2.0 \times 10^{-5} \, {\rm{sec.}} 
                              \bigg( \frac{\langle \hat{S} \rangle}{10^{10} \, \rm{GeV}} \bigg)^2
                              \left( \frac{m_{\widetilde{\tau}}}{100 \, \rm{GeV}} \right)^{-1}.
\end{eqnarray}
In the \textit{2nd. process}, since staus are produced by gauginos which are the decay product of the modulus, they are so abundant that the annihilation becomes effective. 
Thus, the axinos produced in the \textit{2nd. process} can be negligible.  
Comparing the decay width of the stau with the interaction rate by using eq.\eqref{stau's annihilation cross section} and eq.\eqref{stau's decay width}, we can find that  the annihilation of the stau in the $\textit{4th. process}$ does not become effective. 
Thus, when the stau is NLSP the axino abundance is also equal to the gravitino abundance, 
\begin{eqnarray}  \label{stau's yield}
    Y^{\widetilde{\tau}}_{\widetilde{a}} \simeq Y^X_{3/2} 
                 \simeq 4.3 \times 10^{-9} \left( \frac{g_*(T^X_R)}{10} \right)^{-1/4}
                        d_g \left( \frac{N_G}{12} \right)^{1/2}
                        \left( \frac{m_X}{10^6  \, \rm{GeV}} \right)^{1/2}.
\end{eqnarray}

\subsection{Stop and Wino NLSP cases}

We also mention the case of the stop and the wino NLSP briefly. Let us discuss first the case where the stop is the NLSP. If the stop mass is heavy enough to decay into the top quark, the decay width of the stop is obtained by (analogous to eq.\eqref{stau's coupling})
\begin{eqnarray}  \label{stop's decay width}
   \Gamma(\widetilde{t} \to \widetilde{a} t) \simeq
        \frac{1}{32 \pi} \bigg( 1 - \frac{m^2_t}{m^2_{\widetilde{t}}} \bigg)^2
        \left( \frac{m_t}{m_{\widetilde{t}}} \right)^2
        \frac{m_{\widetilde{t}}^3}{\langle \hat{S} \rangle^2},
\end{eqnarray}
and hence the lifetime of the stop is 
\begin{eqnarray}  \label{stop's lifetime}
   \tau_{\widetilde{t}} \simeq 1.8 \times 10^{-8} \, {\rm sec.} 
                               \bigg( \frac{\langle \hat{S} \rangle}{10^{10} \, \rm{GeV}} \bigg)^2
                               \left( \frac{m_{\widetilde{t}}}{200 \, \rm{GeV}} \right)^{-1} 
\end{eqnarray}
with $m_t$ = 174 GeV. 
The annihilation cross section of the stop is obtained by replacing the coupling constant of eq.\eqref{stau's annihilation cross section} with the strong coupling,
\begin{eqnarray}  \label{stop's annihilation cross section}
   \langle \sigma_{\rm{ann.}} v_{\rm{rel.}} \rangle_{\widetilde{t}} 
      \simeq \frac{32 \pi \alpha_3^2}{m^2_{\widetilde{t}}}.
\end{eqnarray}
Thus, when $m_{\widetilde{t}} \gtrsim m_t$ GeV, we can find the annihilation process is not effective in the $\textit{4th. process}$. Then, the abundance of the axino is the same as that of the other NLSP cases,  $i.e.$ eq.\eqref{axino abundance from direct decay of bino}   
\footnote{On the other hand, if $m_{\widetilde{t}} \lesssim m_t$, the decay width of the stop is more suppressed than eq.\eqref{stop's decay width} by the three-body phase space. In such a case, since the annihilation may become  effective, the axino abundance in the $\textit{4th. process}$ is much smaller than that of the case. Therefore, when  $m_{\widetilde{t}} \lesssim m_t$ the axino cannot explain the present DM abundance.}.

If the wino is NLSP 
\footnote{
This does not occur for $c_{\gaugino}=1$ and $\alpha \simeq 1$.
However, once we introduce {\it e.g.} non-universal $c_{\tilde{B}}$ \cite{Abe:2007je}--\cite{Dermisek:2007qi}, 
 it is possible to have wino NLSP, which will not upset the stabilization
 mechanism of $S$ controlled by the strong interaction unless bino is hierarchically heavy.
}, 
its decay width can be obtained by the analogy of that of the bino: 
\begin{subequations}  \label{wino's decay width}
\begin{gather}  
   \Gamma(\widetilde{W} \to \widetilde{a} \gamma) \simeq 
      \frac{\alpha^2_{\rm em} N^2}{256 \pi^3} \frac{1}{\sin^2 \theta_W} 
      \frac{m_{\widetilde{W}}^3}{\langle \hat{S} \rangle^2} \\ 
   \Gamma(\widetilde{W} \to \widetilde{a} Z^0) \simeq 
      \frac{\alpha^2_{\rm em} N^2}{256 \pi^3} \frac{\cos^2 \theta_W}{\sin^4 \theta_W} 
      (1 - x_{Z} ) \left( 1 - \frac{x_{Z}}{2} - \frac{x_{Z}^2}{2} \right) 
      \frac{m_{\widetilde{W}}^3}{\langle \hat{S} \rangle^2},
\end{gather}
\end{subequations}
where $m_{\widetilde{W}}$ is the wino mass. 
For $m_{\widetilde{W}}$ = 100 GeV, the lifetime of the wino is 
\begin{eqnarray}  \label{wino's lifetime}
   \tau_{\widetilde{W}} &=& \Gamma_{\widetilde{W}}^{-1} 
                        \equiv \left( 
                                  \Gamma(\widetilde{W} \to \widetilde{a} \gamma) 
                                + \Gamma(\widetilde{W} \to \widetilde{a} Z^0)
                               \right)^{-1} \nonumber \\
                        &\simeq& 1.9 \times 10^{-3} \, \rm sec.. 
\end{eqnarray}
On the other hand, it is known that the annihilation of the wino pair is very effective. The annihilation cross section is given by \cite{Olive:1989jg}
\begin{eqnarray}  \label{wino's annihilation cross section}
    \langle \sigma_{\rm{ann.}} v_{\rm{rel.}} \rangle_{\widetilde{W}} 
      =  \frac{8 \pi \alpha^2_2}{m^2_{\widetilde{W}}} \frac{(1-x_W)^{3/2}}{(2-x_W)^2}.
\end{eqnarray}
Then we can find that when $m_{\widetilde{W}} \simeq$ 100 GeV, $\langle \sigma_{\rm{ann.}} v_{\rm{rel.}} \rangle_{\widetilde{W}} \, n_{\widetilde{W}} \gg \Gamma_{\widetilde{W}}$ at the gravitino decay, where $n_{\widetilde{W}}$ is the number density of the wino. Thus the abundance of the axino is the same as that of the wino after annihilation,
\begin{eqnarray}
   Y^{\widetilde{W}}_{\widetilde{a}} \simeq \frac{n_{\widetilde{W}}}{s} \bigg|_{T_{3/2}} 
     \simeq 3.3 \times 10^{-11} \left( \frac{g_*(T_{3/2})}{10} \right)^{-1/4}
                                 \left( \frac{m_{3/2}}{10^5 \, \rm GeV} \right)^{-3/2}
                                 \left( \frac{m_{\widetilde{W}}}{100 \, \rm GeV} \right)^2,
\end{eqnarray}
where we have used $x_W \simeq 0.64$ evaluated by $m_{\widetilde{W}}$ = 100 GeV. 
On the other hand, when $m_{\widetilde{W}} \gtrsim$ 300 GeV, the annihilation process does not occur effectively, and then the axino abundance is equal to that of the other NLSP cases. 

Here, we summarize the axino abundance for each NLSP cases in table.1, where we showed only the \textit{4th. process} because the other processes are subdominant. 
When $\widetilde{B}$, $\widetilde{h}$ and $\widetilde{\tau}$ are NLSP cases, the annihilation is always less effective than the decay.  
Since the annihilation cross section of the stop is quite large, the axino abundance becomes too small to explain the present DM abundance when the annihilation of the stop is effective. 

\begin{table}[t]  
  \label{table: abundance of axino} 
   \begin{center} 
    \renewcommand{\arraystretch}{1.5}
    \begin{tabular}{|c|c|c|}
      \hline 
      NLSP & Case A: $\Gamma \gg n \langle \sigma v \rangle$ & Case B: $\Gamma \ll n \langle \sigma v \rangle$ \\  \hline \hline
      $\widetilde{B}$     &                       &                   \\  \cline{1-1}
      $\widetilde{h}$     &                       &   does not occur  \\  \cline{1-1} 
      $\widetilde{\tau}$     & $Y_{\widetilde{a}} \simeq 4.3 \times 10^{-9}$  &     \\  \cline{1-1} \cline{3-3}
      $\widetilde{t}$     &                       & too small abundance  \\  \cline{1-1} \cline{3-3}
      $\widetilde{W}$     &                       & $Y_{\widetilde{a}} \simeq 3.3 \times 10^{-11}$       \\  \hline \cline{1-1} \cline{3-3}
    \end{tabular} 
  \end{center}
  \caption{Table of abundance of the axino in the \textit{4th. process}. Contributions of the other processes are subdominant. Cases A and B, respectively, correspond to situations where the annihilation process of NLSP is not effective and effective. }
\end{table}

\section{Results}  \label{results}

\begin{figure}[t]  
\!\!\!\!\!\!\!\!\!\!\!\!\!\!\!\!\!\!\!\!
 \begin{center} 
   \includegraphics[width=9.0cm, clip]{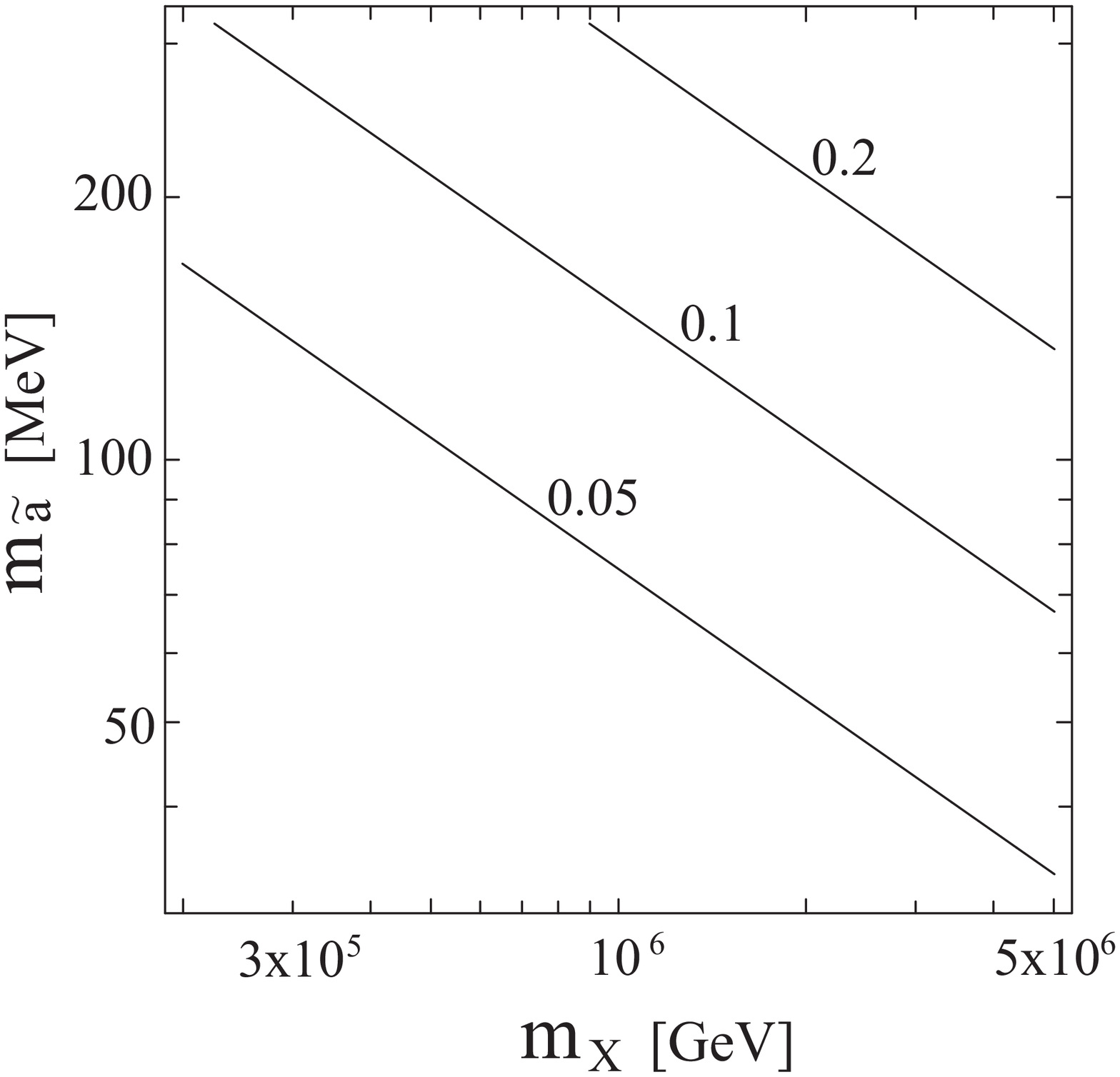} 
   \caption{Contours of the density parameter of the axino, $\Omega_{\widetilde{a}} h^2$, drawn in $m_X- m_{\widetilde{a}}$ plane. Three lines represent $\Omega_{\widetilde{a}} h^2 =0.2, 0.1 $ and $0.05 $, from the above. The contours are the same in any NLSP cases if the decay of NLSP produced by the gravitino decay is more effective than the annihilation process. } 
\label{fig: contour of axion with modulus mass}
 \end{center}
\end{figure}

\begin{figure}[b]  
   \begin{center}
   \includegraphics[width=9cm, clip]{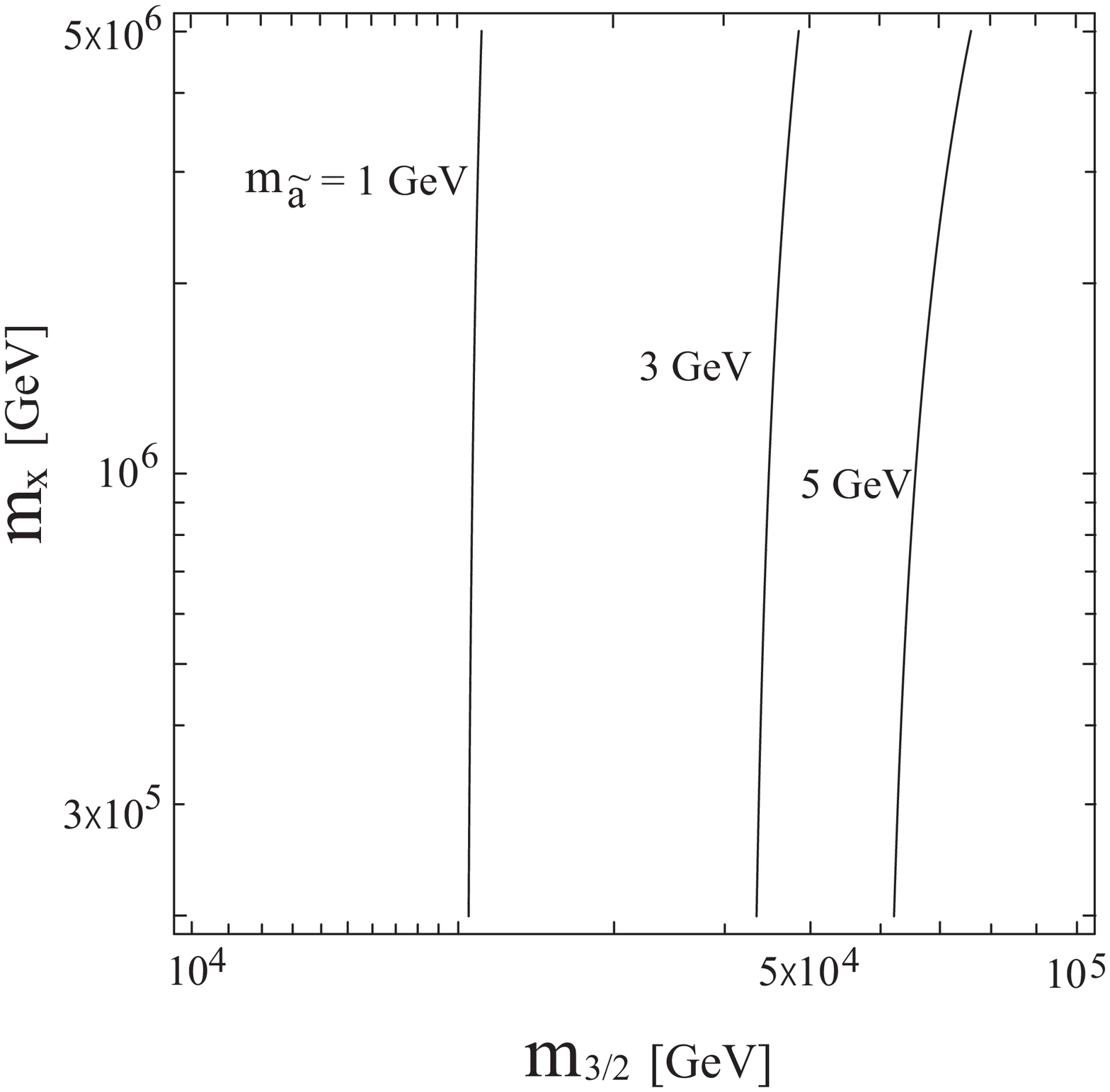} 
   \caption{Contours of the axino mass, $m_{\widetilde{a}}$, which satisfy $\Omega_{\widetilde{a}} h^2 = 0.1$ drawn in $m_{3/2}- m_{X}$ plane in the case where NLSP is the wino with $m_{\widetilde{W}} \simeq 100$ GeV and their annihilation is effective. Three lines represent $m_{\widetilde{a}} =1, 3 $ and 5 GeV, from left to right. Here, we set $\lambda = N = 1$, $k = 0$ and $m_{\widetilde{W}}$ = 100 GeV for simplicity.  From the gravitino problem point of view, the gravitino mass should be heavier than about 30 TeV.  } 
\label{fig: contour of axino in wino NLSP case}
   \end{center}
\end{figure}

At last, we are now ready to discuss whether the axino relic abundance can explain the present DM abundance. 
If the decay of NLSP is more effective than the annihilation in the $\textit{4th. process}$, the $\Omega$ parameter, which is defined by the ratio of the mass density of $\widetilde{a}$ and the critical density, is the same in any NLSP cases
\begin{eqnarray}   \label{omega of axino for bino NLSP}
   \Omega_{\widetilde{a}} h^2
                        \simeq 1.2 \left( \frac{m_{\widetilde{a}}}{1 \, \rm{GeV}} \right)
                        \left( \frac{g_*(T^X_R)}{10} \right)^{-1/4}
                        d_g \left( \frac{N_G}{12} \right)^{1/2}
                        \left( \frac{m_X}{10^6  \, \rm{GeV}} \right)^{1/2}.
\end{eqnarray} 
On the other hand, if the wino is NLSP and their annihilation is effective, the $\Omega$ parameter is
\begin{eqnarray}  \label{omega of axino for wino NLSP}
   \Omega_{\widetilde{a}} h^2 \simeq 10^{-3} \left( \frac{m_{\widetilde{a}}}{1 \, \rm GeV} \right) 
                          \!\!\!\! &\bigg\{& \!\!\!\!
                                        26 \left( \frac{g_*(T_{3/2})}{10} \right)^{-1/4}
                                            \left( \frac{m_{3/2}}{50 \, \rm TeV} \right)^{-3/2}
                                            \left( \frac{m_{\widetilde{W}}}{100 \, \rm GeV} \right)^2 \nonumber \\
                                     &&+ \Big( 4.8 + 1.5 \lambda^4 N^2 k^2 \Big) 
                                            \left( \frac{g_*(T_R^X)}{10} \right)^{-1/4}
                                            \left( \frac{m_X}{10^6 \, \rm GeV} \right)^{1/2}
                                    \bigg\}.
\end{eqnarray}
Here we have set $d_g = 1$ and $N_G = 12$. 
In fig.\ref{fig: contour of axion with modulus mass}, we plotted eq.\eqref{omega of axino for bino NLSP} in terms of $m_X$ and $m_{\widetilde{a}}$. 
From the WMAP three year results \cite{Spergel:2006hy}, the DM abundance in the present Universe is $\Omega_{\rm{DM}} h^2 = 0.105^{+0.007}_{-0.013}$ (68 $\%$ C.L.). 
Thus,  we can find from fig.\ref{fig: contour of axion with modulus mass} that the axino with $m_{\widetilde{a}} \simeq \mathcal{O}(100) $ MeV can explain the present DM abundance in any NLSP cases, if the decay of NLSP is more effective than the annihilation in the $\textit{4th. process}$. 
In fig.\ref{fig: contour of axino in wino NLSP case}, we have plotted axino mass contours which satisfy $\Omega_{\widetilde{a}} h^2 = 0.1$ in $m_{3/2}-m_X$ plane for the wino NLSP case, where we set $\lambda = N =1$, $k = 0$, and $m_{\widetilde{W}} =$ 100 GeV.  
If the wino is NLSP and their annihilation is effective, fig.\ref{fig: contour of axino in wino NLSP case} shows that the right amount of DM can be explained by the axino with a few GeV mass 
\footnote{Here, we should consider that the gravitino mass is heavier than about 30 TeV not to spoil the success of BBN \cite{Kawasaki:2004yh}--\cite{Kohri:2005wn}. }. 
Here we should caution that the estimate of the relic abundance given in this paper is rather rough, which may contain an error of factor 2 or so. 
From eq.\eqref{eq: mass of singlino}, we can parametrize the axino mass by using $\lambda$ as 
\begin{eqnarray}  \label{parameterization of axino mass}
   m_{\widetilde{a}} = \frac{\alpha_3 \lambda^2}{4 \pi^3} {F_\Phi}.
\end{eqnarray} 
Substituting eq.\eqref{parameterization of axino mass} into $\Omega$ parameters, we can find 
\begin{eqnarray}  \label{omega of axino with lambda}
   \Omega_{\widetilde{a}} h^2 \simeq 0.1 \, \left( \frac{\lambda}{0.06} \right)^2  
                                 \left( \frac{{F_\Phi}}{50 \, \rm{TeV}} \right)
                                 \left( \frac{m_X}{10^6  \, \rm{GeV}} \right)^{1/2}
                                 \left( \frac{g_*(T^X_R)}{10} \right)^{-1/4}
                                 d_g \left( \frac{N_G}{12} \right)^{1/2},
\end{eqnarray}
for the case where the annihilation is not effective, and 
\begin{eqnarray}  \label{omega of axino with lambda in wino case}
   \Omega_{\widetilde{a}} h^2 \simeq \left( \frac{\lambda}{0.4} \right)^2 
                                     \left( \frac{{F_\Phi}}{50 \, \rm{TeV}} \right)
                             \!\!\! &\bigg\{& \!\!\!
                                        0.09 \left( \frac{g_*(T_{3/2})}{10} \right)^{-1/4}
                                             \left( \frac{m_{3/2}}{50 \, \rm TeV} \right)^{-3/2}
                                             \left( \frac{m_{\widetilde{W}}}{100 \, \rm GeV} \right)^2 \nonumber \\
                                 &&+ \, 0.02 \left( \frac{g_*(T_R^X)}{10} \right)^{-1/4}
                                             \left( \frac{m_X}{10^6 \, \rm GeV} \right)^{1/2}
                                     \bigg\},
\end{eqnarray}
for the wino NLSP case, whose annihilation process is effective. 
Thus, $\mathcal{O}(0.1)$ coupling between the axion superfield and messengers leads to the right amount of DM in any NLSP cases.

Here, we should mention the free-streaming scale of the axino LSP.  
According to the discussion in the previous section, the axino LSPs are produced mainly at the decays of non-relativistic NLSPs. At the production, they are relativistic and travel freely until their momentum gets red-shifted to be non-relativistic. 
The observations of Lyman-$\alpha$ forest require that the free-streaming scale of DM have to be less than $\mathcal{O}(1)$ Mpc \cite{Viel:2005qj} \cite{Narayanan:2000tp}. 
We can estimate the free-streaming scale of the axino as \cite{Borgani:1996ag} \cite{Cembranos:2005us}
\begin{eqnarray}  \label{free-streaming}
   \lambda_{\rm FS} = \int^{t_{\rm EQ}}_{t_{\widetilde{a}}} \frac{v(t)}{a(t)} dt
                    \simeq 1.0 \, {\rm Mpc} \, u_{\widetilde{a}} 
                           \left[ \frac{t_{\widetilde{a}}}{10^6 \, \rm s} \right]^{1/2} 
                           \left\{ 
                              1 + 0.14 \ln 
                                 \bigg[ \bigg( 
                                           \frac{10^6 \, \rm s}{t_{\widetilde{a}}} 
                                        \bigg)^{1/2} \frac{1}{u_{\widetilde{a}}} 
                                 \bigg]
                           \right\},
\end{eqnarray}
where $v(t)$ and $a(t)$ are the velocity of the axino and the scale factor, respectively. 
$t_{\rm EQ}$ is the time at matter-radiation equality, $t_{\widetilde{a}}$ the time when the axino is produced and  
the three-momentum normalized by the axino mass  $u_{\widetilde{a}} = p/ m_{\widetilde{a}}$. 
From the discussion of section \ref{cosmology}, we can approximate that $t_{\widetilde{a}}$ is equal to the lifetime of the gravitino, and hence we find 
\begin{eqnarray}  \label{free-streaming of axino}
   \lambda_{\rm FS} \simeq 0.4 \, {\rm Mpc} \left(\frac{m_{\rm NLSP}}{200 \, \rm GeV} \right)
                           \left( \frac{m_{\widetilde{a}}}{0.1 \, \rm GeV} \right)^{-1}
                           \left( \frac{m_{3/2}}{50 \, \rm TeV} \right)^{-3/2}, 
\end{eqnarray}
for any NLSP cases if the annihilation process is not effective. The axinos produced directly by the decays of moduli and gravitinos have much larger free-streaming length of order 10 Mpc. However, their contamination in the total dark matter density is about 1$\%$ level, so that their contribution will be harmless \cite{Viel:2005qj}. 

On the other hand, in the case where the annihilation process is effective and NLSP is the wino with $m_{\widetilde{W}} = 100$ GeV, axinos produced by gravitinos and the modulus have the free-streaming scale of the order of $\mathcal{O}(1)$ Mpc 
\footnote{The free-streaming scale of the axino produced by the stau or the wino NLSP is $\mathcal{O}(0.01)$ Mpc.}:
\begin{subequations}  \label{FS axino for wino NLSP}
\begin{gather}
   \lambda_{\rm FS} \simeq 1.3 \, {\rm Mpc} 
                           \left(\frac{m_{3/2}}{10^5 \, \rm GeV} \right)^{-1/2}
                           \left( \frac{m_{\widetilde{a}}}{5 \, \rm GeV} \right)^{-1}, \\ 
   \lambda_{\rm FS} \simeq 0.4 \, {\rm Mpc}     
                           \left(\frac{m_{X}}{5 \times 10^6 \, \rm GeV} \right)^{-1/2}
                           \left( \frac{m_{\widetilde{a}}}{5 \, \rm GeV} \right)^{-1}.                   
\end{gather}
\end{subequations}
The free-streaming scales of the axino, eq.\eqref{free-streaming of axino} and eq.\eqref{FS axino for wino NLSP}, are in an interesting range for the small structure problem point of view. The model of the cold dark matter (CDM) can explain the large scale structure of the Universe very well. However,  $N$-body simulations in the CDM model seem to conflict with observations at the smaller scale ($\lesssim$ Mpc). Numerical simulations based on the CDM model predict that the number of halos is by one order of magnitude larger than observations within $\mathcal{O}(1)$ Mpc. 
This is called the missing satellite problem \cite{Klypin:1999uc}. 
Another possible discrepancy is that the prediction of simulations for mass profile of CDM halos is excessively cuspy compared with observations, which is known as the cusp problem \cite{Flores:1994gz}. 
One of the resolutions to these problems is to invoke some DM whose free-streaming scale is in the range of $\mathcal{O} (0.1 - 1)$ Mpc, because it can sweep out density fluctuations in the small scale \cite{Lin:2000qq}. 
Then the axino LSP of our model may also be able to solve these problems.

Before concluding this paper, we also mention whether the axion can explain DM. 
We have already discussed that the singlet, $S$, can be stabilized at $10^{10}$ GeV to $10^{12}$ GeV by choosing the modular weight of the messenger appropriately. The axion relic abundance eq.\eqref{axion abundance} with $\Theta = \mathcal{O}(1)$ shows that when the PQ scale is obtained $f_{\rm{PQ}} \simeq 10^{12}$ GeV the axion can constitute DM abundance.  
When the axion superfield is stabilized at $10^{12}$ GeV, however, the lifetime of the saxion is longer than that of the modulus, 
then the Universe may be dominated by the saxion if the initial amplitude of the saxion is not so small compared with the Planck scale. 
Since the saxion mainly decays into axions, it will spoil the success of BBN.  
Therefore, to explain DM by the axion would be difficult in our model unless the initial amplitude of the saxion is suppressed or there is an extra entropy production.

\section{Summary}  \label{Summary}
Let us summarize this paper. We have discussed the axionic extended mirage mediation, \textit{axionic mirage mediation}, to remedy crucial problems included in the mirage mediation: the $\mu$-problem and the moduli problem. 
We showed that the PQ scale can be obtained in the axion window, $10^9 ~{\rm GeV} \lesssim f_{\rm PQ} \lesssim 10^{12 - 13}$ GeV, by stabilizing the axion superfield. 
Especially, in the KKLT set-up ($\alpha \simeq 1$) with $\ell$ = 1/2, we obtained the PQ scale at $10^{10}$ GeV.  
To obtain desired values of $\mu$-/$B \mu$-term was achieved by introducing some singlets with couplings eq.\eqref{mu-term superpotential} and eq.\eqref{mu-term f function} and integrating out the singlet $T$ which becomes heavy due to the VEV of the axion superfield.   
It was also found that the phase of the $B$-parameter can be rotated away simultaneously with the phases of the gaugino mass and the $A$-parameter. Thus, SUSY CP problem is also absent in our model.

We investigated the implications of the modulus, whose mass is larger than that of the gravitino by several order of magnitude, and the saxion to cosmology. 
When both the saxion and the modulus have the order of the Planck scale initial amplitude, the Universe is dominated by the oscillation energy of the modulus eventually. 
In this case, we found that as long as the saxion lifetime is sufficiently shorter than that of the modulus, axions produced by the saxion decay would not spoil the success of BBN.  
Thus, we discussed the cosmic evolution only after the modulus decay. 
Although there were several axino LSP production processes, the dominant contribution to the axino relic abundance came from the decay of the NLSPs produced by the gravitinos. 
We have estimated the axino abundance for the various NLSP cases. However,  we found  that the axino relic abundance is the same in any NLSP cases, if the decay width of NLSP into the axino is larger than their annihilation rate. In such a situation, we found that the axino mass with $\mathcal{O}(100)$ MeV  leads to the present DM abundance. 
On the other hand, in the case where NLSP is the wino and their interaction rate is more effective than their decay process, the axino abundance is less than the former case about two orders. In this case, the DM abundance can be explained by the axino with $m_{\widetilde{a}} \simeq \mathcal{O}(1)$ GeV. 
When we parametrize the axino mass in terms of the Yukawa coupling of $S$, $\lambda$, by eq.\eqref{parameterization of axino mass}, $\lambda \simeq \mathcal{O}(0.1)$ explains the right amount of DM abundance, eq.\eqref{omega of axino with lambda} and eq.\eqref{omega of axino with lambda in wino case}.  
Thus, nevertheless the branching ratio of the modulus to the gravitino pair is sizable, our model can solve the moduli problem as well as leads to right amount of DM abundance.
In addition, we also found that the free-streaming scale of the axino LSP is in an interesting region $\mathcal{O}(0.1 - 1)$ Mpc for the small structure problems.

Finally we would like to briefly mention signatures of neutralino NLSP decays at collider experiments.
As was discussed in this paper, the axino LSP couples to the Higgs multiplets at tree level, whereas 
its coupling to the gauge sector arises at one loop, and thus is suppressed. This implies that the 
neutralino NLSP dominantly decays to the Higgs boson (as well as the $Z^0$ boson) as far as it contains a sizable fraction of the higgsino components.  We expect that the NLSP decay into the Higgs boson will provide a spectacular signal of displaced vertex emitting hard jets, if the decay vertex can be 
reconstructed. Detailed study on the collider signatures will be presented elsewhere.

\section*{{Acknowledgement}}
We would like to thank T. Higaki and A. Yotsuyanagi for useful discussions. 
The work was partially supported by the grants-in-aid from the Ministry of Education, Science, Sports, and Culture of Japan, No.16081202 and No.17340062. 
K.O. is supported by the Grand-in-aid for Scientific Research No.19740144
from the Ministry of Education, Culture,
Sports, Science and Technology of Japan.
K.O. also thanks Yukawa Institute in Kyoto University for the use of
Altix3700 BX2.

\end{document}